\documentclass[aps,floatfix,twocolumn,floats,superscriptaddress]{revtex4-1}

\usepackage[T1]{fontenc}
\usepackage{times}
\usepackage{graphicx}
\usepackage{amsmath}
\usepackage{amsfonts}
\usepackage{amssymb}
\usepackage{color}
\usepackage{braket}
\usepackage{textcomp}
\usepackage{comment}
\usepackage[normalem]{ulem}

\begin{document}

\title{Variational principles in quantum Monte Carlo: the troubled story of variance minimization}

\author{Alice Cuzzocrea}
\affiliation{MESA+ Institute for Nanotechnology, University of Twente, P.O. Box 217, 7500 AE Enschede, The Netherlands}
\author{Anthony Scemama}
\affiliation{Laboratoire de Chimie et Physique Quantiques, Universit\'e de Toulouse, CNRS, UPS, France}
\author{Wim J. Briels}
\email{w.j.briels@utwente.nl}
\affiliation{MESA+ Institute for Nanotechnology, University of Twente, P.O. Box 217, 7500 AE Enschede, The Netherlands}
\author{Saverio Moroni}
\email{moroni@democritos.it}
\affiliation{CNR-IOM DEMOCRITOS, Istituto Officina dei Materiali, and SISSA Scuola Internazionale Superiore di Studi Avanzati, Via Bonomea 265, I-34136 Trieste, Italy}
\author{Claudia Filippi}
\email{c.filippi@utwente.nl}
\affiliation{MESA+ Institute for Nanotechnology, University of Twente, P.O. Box 217, 7500 AE Enschede, The Netherlands}

\begin{abstract}
We investigate
the use of different variational principles in quantum Monte Carlo, namely energy and variance
minimization, prompted by the interest in the robust and accurate estimate of electronic excited states.
For two prototypical, challenging molecules, we readily reach the accuracy of the best available reference excitation energies
using energy minimization in a state-specific or state-average fashion for
states of different or equal symmetry, respectively.
On the other hand, in variance minimization, where the use of suitable functionals is expected 
to target specific states regardless of the symmetry, we encounter severe problems for a variety of wave
functions: as the variance converges, the energy drifts away from that of the selected state.
This unexpected behavior is sometimes observed even when the target is the ground state,
and generally prevents the robust estimate of total and excitation energies.
We analyze this problem using a very simple wave function and infer that
the optimization finds little or no barrier to escape from a local minimum or 
local plateau, eventually converging to the unique lowest-variance state instead of the target state.
While the loss of the state of interest can be delayed and possibly avoided by reducing the statistical 
error of the gradient, for the full optimization of realistic wave functions, variance minimization with current functionals appears to be an impractical
route.

\end{abstract}

\maketitle

\section{Introduction}
\label{sec:intro}

Light-induced processes are at the heart of a variety of phenomena and applications which range
from harnessing the response to light of biological systems to improving the technologies for renewable
energies. The contribution of electronic structure theory in this field hinges on its ability to efficiently 
and accurately compute excited-state properties.  
In this context, the use of quantum Monte Carlo (QMC) methods is relatively recent and quite 
promising~\cite{filippi2009,zimmerman2009,valsson2010,send2011,valsson2013,guareschi2016,hunt2018,blunt2019,dash2019}:
QMC approaches provide an accurate (stochastic) solution of the Schr\"odinger 
equation and benefit from a favorable scaling with system size and great ease of 
parallelization~\cite{Foulkes2001,luchow_quantum_2011,austin_quantum_2012}.  
Importantly, recent methodological advancements~\cite{sorella2010,neuscamman2012,filippi2016,assaraf2017} 
enable the fast calculation of energy derivatives and the optimization of many thousands of 
parameters for the internally consistent computation of QMC wave functions and geometries in the ground
and excited states~\cite{dash2018,dash2019}.

Here, we investigate the use of two different variational principles for ground and excited states 
in QMC, namely, variance and energy minimization, to assess whether they allow us 
to fully capitalize on the increased power of minimization algorithms and availability of
accurate wave functions.
Variance minimization techniques~\cite{coldwell1977,umrigar1988,malatesta1997,kent1999,umrigar2005} have been extensively
employed in QMC for the last 30 years but their potential 
for the computation of excited states has only recently been revisited and exploited to compute 
vertical excitation energies of various small molecules~\cite{shea2017,pineda2019}. Different 
functionals for the optimization of the variance~\cite{umrigar1988,umrigar2005,shea2017}
have also been put forward with the common attractive feature of the built-in possibility 
to target a specific state and avoid in 
principle the complications encountered in energy minimization where, without constraints, 
one would generally collapse to lower-energy states. 

For our study, we select two molecules, a small cyanine dye and a retinal model, because of the 
difficulties they pose in the computation of the lowest vertical excitation 
energy~\cite{send2011,valsson2012,huix2013,tuna2015,Guennic2015}, and the different requirements
in the procedure adopted in energy minimization: while the ground and excited states of the cyanine
belong to different symmetries  and can therefore be treated in a state-specific manner, this is not
the case for the retinal model, where energy minimization must be performed in a state-average fashion. 
For both molecules and therefore regardless of the nature of the optimization, we find that
energy minimization leads to the stable and fast convergence of the total energies of
the states of interest. 
Furthermore, with the use of compact and balanced energy-minimized wave functions constructed through a selected 
configuration interaction (CI) approach, we recover vertical excitation energies 
which are already at the variational Monte Carlo (VMC) level within chemical accuracy (about 0.04 eV) 
of the reference coupled cluster or extrapolated CI values.
On the other hand, for both molecules and for nearly all wave functions investigated, the optimization of 
all parameters in variance minimization is problematic since it results in the apparent loss of the state of 
interest over sufficiently long optimization runs, precluding the estimate of the excitation energy.  
This occurs for the different functionals originally proposed 
to stabilize the optimization and, surprisingly, in some cases also when targeting the ground state.  This finding is 
unexpected, especially considering that variance minimization has been the method of choice in QMC 
for decades and is still routinely used, albeit for wave functions with a small number of parameters or 
where the optimization is limited either to few optimization steps or to the Jastrow factor.

To understand these newly-found issues, we examine how variance minimization behaves when optimizing the linear
coefficients of a very simple wave function.  Working in the linear sub-space spanned by a few approximate
eigenvectors, 
we discover that the optimization of the CI parameters in variance minimization does not converge to the target 
eigenstate but to a different one: for approximate wave functions, the multiple minima of the variance have generally
different values and, during the minimization, the system slowly reaches the eigenstate corresponding 
to the absolute minimum of the variance, no matter what the starting state is.
We show that a similar 
behavior can be inferred also for more complicated wave functions and, while the process can be slowed down by 
reducing the statistical error on the gradient of the variance driving the minimization,
it severely limits the use of variance minimization for the optimization of realistic wave functions.

In Section~\ref{sec:methods}, we recap the equations used for energy and variance optimization, 
discuss the procedure employed for the state-average case, and introduce the 
ingredients for a stable version of the Newton method in variance minimization. 
In Section~\ref{sec:comput}, we summarize the computational details and, in Section~\ref{sec:results},
present the accurate vertical excitation energies obtained in energy minimization 
and the difficulties encountered in variance minimization for both molecules.
We elucidate these findings and conclude in Section~\ref{sec:disc}.

\section{Methods}
\label{sec:methods}

We briefly introduce below the variance and energy minimization approaches used to optimize the wave
functions in variational Monte Carlo. While we employ variance minimization as a state-specific approach
to target a given state, we must distinguish between a state-specific and a state-average route for energy
optimization when the excited state of interest is of different or equal symmetry, respectively,
than other lower-lying states.

\subsection{Wave function form}

The wave functions employed in this work are of the Jastrow-Slater type, namely, the product of a
determinantal expansion and a Jastrow correlation function, $\mathcal{J}$, as
\begin{eqnarray}
\Psi = \mathcal{J} \sum_{i=1}^{N_{\rm det}} c_i D_i\,,
\end{eqnarray}
where the determinants are expressed on single-particle orbitals and the Jastrow factor includes an
explicit dependence on the electron-electron distances. Here, the Jastrow factor is chosen to include
electron-electron and electron-nucleus correlation terms~\cite{Jastrow}.  For the determinantal
component, we select the relevant determinants according to different recipes: i) very simple ansatzes 
such as Hartee-Fock (HF) or a CI singles (CIS) expansion recently put forward as a computationally cheap and sufficiently 
accurate wave function for excited states in QMC~\cite{neuscamman2016,blunt2019}; ii) 
complete-active-space (CAS) expansions where small sets of important active orbitals are manually identified; 
iii) CI perturbatively selected iteratively (CIPSI) expansions generated to yield automatically balanced multiple states.
All expansions are expressed in terms of spin-adapted configuration state functions (CSF) 
to reduce the number of variational parameters.

\subsection{Energy minimization}

For state-specific optimization in energy minimization, we employ the stochastic reconfiguration
(SR) method~\cite{sorella2007,neuscamman2012} in a low-memory conjugate-gradient implementation~\cite{neuscamman2012}.
Given a starting wave function $\Psi$ depending on a set of parameters $\textbf{p}$, we denote the 
derivatives of $\Psi$ with respect to a parameter $p_i$ as $\Psi_i = \partial_i \Psi$.
At every step of the SR optimization, the parameter variations, $\Delta p_i$, are computed according to 
the equation:
\begin{equation}
\bar{\mathbf{S}}\Delta \textbf{p} = -\frac{\tau}{2}\mathbf{g}\,,
\label{eq:sr}
\end{equation}
where $\tau$ is a positive quantity chosen small enough to guarantee the convergence. The vector $\mathbf{g}$ 
is the gradient of the energy with components:
\begin{eqnarray}
g_i = \frac{\partial E}{\partial p_i} 
&=& 2 \left[\frac{\braket{\Psi_i|\mathcal{\hat{H}}|\Psi}}{\braket{\Psi|\Psi}} - E \frac{\braket{\Psi|\Psi_i}}{\braket{\Psi|\Psi}} \right] \nonumber\\
&=& 2 \left[\left\langle\frac{\Psi_i}{\Psi}E_{\rm L}\right\rangle-\langle E_{\rm L}\rangle\left\langle\frac{\Psi_i}{\Psi}\right\rangle\right]\,,
\label{eq:gradsr}
\end{eqnarray} 
where $E_{\rm L}= \hat{H}\Psi/\Psi$ is the so-called local energy and $\langle .\rangle$ denotes the 
Monte Carlo average of the quantity in brackets over the electron configurations sampled from 
$\Psi^2/\braket{\Psi|\Psi}$. The matrix $\bar{\mathbf{S}}$ has components:
\begin{eqnarray}
\bar{S}_{ij} &=& \frac{\braket{\Psi_i|\Psi_j}}{\braket{\Psi|\Psi}} - \frac{\braket{\Psi|\Psi_i}}{\braket{\Psi|\Psi}}\frac{\braket{\Psi|\Psi_j}}{\braket{\Psi|\Psi}} \nonumber\\
             &=& \left\langle\frac{\Psi_i}{\Psi}\frac{\Psi_j}{\Psi}\right\rangle - \left\langle\frac{\Psi_i}{\Psi}\right\rangle\left\langle\frac{\Psi_j}{\Psi}\right\rangle \equiv \left\langle\frac{\bar{\Psi}_i}{\Psi}\frac{\bar{\Psi}_j}{\Psi}\right\rangle\,,
\label{eq:s-sr}
\end{eqnarray} 
which is expressed in the last equality as the overlap matrix in the semi-orthogonal basis, 
$\bar{\Psi}_i=\Psi_i -[\braket{\Psi|\Psi_i}/\braket{\Psi|\Psi}]\Psi$.

When the state of interest is energetically not the lowest in its symmetry class,  we start from a set of wave functions for
the multiple states which share the same Jastrow factor and orbitals but are characterized by different linear CI 
coefficients as
\begin{eqnarray}
\Psi^{I} = \mathcal{J} \sum_{i=1}^{N_{\rm det}} c_i^{I} D_i \,,
\end{eqnarray}
where the superscript $I$ indicates a particular state. To obtain a balanced description of the states of interest, we 
optimize the non-linear parameters of the orbitals and the Jastrow factor by minimizing the state-average energy~\cite{filippi2009}:
\begin{eqnarray}
E^{\rm SA} = \sum_{I} w_{I} \frac{\braket{\Psi^{I}|\mathcal{\hat{H}}|\Psi^{I}}}{\braket{\Psi^{I}|\Psi^{I}}}\,,
\end{eqnarray}
where the weights $w_{I}$ are kept fixed and $\sum_I w_I=1$.  
To this aim, we follow the SR scheme (Eq.~\ref{eq:sr}) and use the gradient of the state-average energy
\begin{eqnarray}
g_i^{\rm SA} = \sum_{I} w_{I} g_i^{I}\,,
\end{eqnarray}
where $g_i^{I}$ is the gradient with respect to a parameter $p_i$ of the energy of state $I$, which is computed 
from the wave function $\Psi^I$ and its derivatives as in Eq.~\ref{eq:gradsr}.  Moreover, in analogy to the single-state
optimization, we introduce a weighted-average overlap matrix defined as 
\begin{eqnarray}
\bar{S}_{ij}^{\rm SA} = \sum_{I} w_{I} \bar{S}^{I}_{ij}\,,
\end{eqnarray}
where the overlap matrix for each state is computed from the corresponding wave function as in Eq.~\ref{eq:s-sr}. We 
stress that, while the state-average SR procedure is defined simply by analogy with the single-state case, it does lead
to the minimization of the state-average energy since we employ the appropriate ${\bf g}^{\rm SA}$.

We alternate a number of optimization steps of the non-linear parameters with the optimization of the linear coefficients 
$c_i^I$, whose optimal values are the solution of the generalized eigenvalue equations
\begin{equation}
{\bf H}^{\rm CI}{\bf c}^I = E_I {\bf S}^{\rm CI} {\bf c}^I\,,
\label{eq:eqci}
\end{equation}
where the Hamiltonian and overlap matrix elements are defined in the basis of the functions $\{\mathcal J D_i\}$ 
and estimated through Monte Carlo sampling.
After diagonalization of Eq.~\ref{eq:eqci}, 
orthogonality between the individual states is automatically enforced.  To solve the eigenvalue equation with 
a memory efficient algorithm, we use the Davidson diagonalization scheme in which the lowest energy eigenvalues 
are computed without the explicit construction of the entire Hamiltonian and overlap matrices~\cite{neuscamman2012}.
A similar procedure was recently followed in Ref.~\cite{Sabzevari2020}.

\subsection{Variance minimization}

To perform variance minimization, we can directly minimize the variance of the state of interest,
\begin{eqnarray}
\sigma^2 = \frac{\braket{\Psi|(\mathcal{\hat{H}}-E)^2|\Psi}}{\braket{\Psi|\Psi}}\,,
\end{eqnarray}
or follow a somewhat more stable optimization procedure by minimizing the expression 
\begin{eqnarray}
\sigma^2_{\omega} = \frac{\braket{\Psi|(\mathcal{\hat{H}}-\omega)^2|\Psi}}{\braket{\Psi|\Psi}}\,,
\end{eqnarray}
where the energy $\omega$ is fixed during the optimization step and then appropriately modified to follow 
the current value of the energy as originally proposed in Ref.~\cite{umrigar1988}.
Recently, a functional $\Omega$ has been put forward,
\begin{eqnarray}
\Omega =  \frac{\braket{\Psi|(\omega - \mathcal{\hat{H}})|\Psi}}{\braket{\Psi|(\omega -\mathcal{\hat{H}})^2|\Psi}}\,,
\end{eqnarray}
whose minimization is equivalent to variance minimization if $\omega$ is eventually updated to the running value
of $E-\sigma$~\cite{shea2017}. 

Because of its simplicity, we choose here the functional $\sigma^2_{\omega}$ but 
also compare the convergence behavior obtained with the functional $\Omega$. 
To this aim, we use the Newton optimization method as in Ref.~\cite{umrigar2005} and update
the parameters as 
\begin{equation}
\Delta \textbf{p} = -\tau \mathbf{h}^{-1} \mathbf{g}\,,
\end{equation}
where $\mathbf{g}$ is here the gradient of $\sigma^2_{\omega}$ and $\mathbf{h}$ its Hessian matrix,
and the parameter $\tau$ is introduced to damp the size of the variations.

The components of the gradient are given by
\begin{eqnarray}
g_i  &=& 2\left[ \left\langle \frac{\mathcal{\hat{H}} \Psi_i}{\Psi}E_{L}\right\rangle 
         -\left\langle\frac{\Psi_i}{\Psi}\right\rangle \langle E_L^2\rangle \right.\\
     &-& \left.\omega\left(\left\langle\frac{\Psi_i}{\Psi}E_{\rm L}\right\rangle 
        +\left\langle\frac{\hat{H}\Psi_i}{\Psi}\right\rangle 
        -2\left\langle\frac{\Psi_i}{\Psi}\right\rangle \langle E_L\rangle \right) \right]\,,\nonumber
\label{eq:gradsigma_o}
\end{eqnarray}
and we discuss other possible equivalent expressions and their relative fluctuations in the SI.
The Hessian matrix elements require the second derivatives of the wave function and, to avoid their computation, 
we follow the same approximation strategy of the Levenberg-Marquardt algorithm~\cite{numrec}
and manipulate the expression of the variance in a somewhat different way than 
proposed in Refs.~\cite{malatesta1997,umrigar2005,toulouse2008},
to obtain the approximate expression of the Hessian matrix
\begin{eqnarray}
\label{eq:h2}
h_{ij}=&&\left\langle \left[\partial_i E_{\rm L}+ (E_{\rm L} -\omega)
                             \left(\frac{\Psi_i}{\Psi} -\left\langle \frac{\Psi_i}{\Psi}\right\rangle\right)\right]\right.\\
       \times &&\left. \,\ \ \left[\partial_j E_{\rm L} + (E_{\rm L} -\omega)
                             \left(\frac{\Psi_j}{\Psi} -\left\langle \frac{\Psi_j}{\Psi}\right\rangle\right)\right]\right\rangle\,,\nonumber
\end{eqnarray} 
Details of the derivation and alternative expressions for
the Hessian are given in the SI. 

We use the Newton method and the Hessian ${\bf h}$ (Eq.~\ref{eq:h2}) when optimizing both $\sigma_\omega^2$ and the $\Omega$ 
functional in combination with the corresponding gradient. Furthermore, we follow Ref.~\cite{shea2017} in keeping $\omega$ fixed 
to an appropriate guess energy for an initial number of minimization steps, upgrading it linearly to the running energy 
(or $E-\sigma$ in the case of $\Omega$) over some intermediate iteration steps, and then setting it equal to the current 
energy estimate for the rest of the run.

\section{Computational details}
\label{sec:comput}

All QMC calculations are carried out with the program package CHAMP~\cite{Champ}. We employ
scalar-relativistic energy-consistent HF pseudopotentials and the correlation-consistent
Gaussian basis sets specifically constructed for these pseudopotentials~\cite{burkatzki2007,BFD_H2013}.
Unless otherwise specified, we use a double-$\zeta$ basis set minimally augmented with $s$ and $p$ 
diffuse functions on the heavy atoms and denoted here as maug-cc-pVDZ.  Basis-set convergence tests 
are performed with the fully augmented double (aug-cc-pvDZ) and triple (aug-cc-pvTZ) basis sets. 
In all cases, the exponents of the diffuse functions are taken from the corresponding all-electron 
Dunning's correlation-consistent basis sets~\cite{kendall1992}.

In the state-specific (energy and variance) optimization runs, we sample a guiding wave function that differs from 
the current wave function close to the nodes~\cite{attaccalite2008} to guarantee finite variances of 
the estimators of the gradient, overlap, and Hessian matrix elements. 
In the state-average energy minimizations, we employ equal
weights for the multiple states and sample a guiding wave function constructed as $\Psi_g^2 =
\sum_I |\Psi^I|^2$, to ensure that the distribution sampled 
has a large overlap with all states of interest~\cite{filippi2009}.
All wave function parameters (Jastrow, orbital, and CI coefficients) are optimized and the damping factor, $\tau$, in the SR and the Newton method
is set to 0.05 and 0.1, respectively, unless otherwise specified. In the DMC calculations,
we treat the pseudopotentials beyond the locality approximation using the T-move algorithm~\cite{casula2006a}
and employ an imaginary time-step of 0.05 a.u.\ which yields excitation energies converged to better
than 0.01 eV as shown in the SI.

The HF, CIS, and complete-active-space self-consistent-field (CAS\-SCF) calculations are carried out
with the program GAMESS(US)~\cite{schmidt1993,gordon2005}. For the cyanine dye, we consider different CAS
expansions: a CAS(6,5) and CAS(6,10) correlating 6 $\pi$ electrons in the orbitals constructed from the
$2p_z$ and $3p_z$ atomic orbitals; a truncated CAS(14,13) consisting of 6 $\pi$ and 8 $\sigma$ electrons 
in 13 bonding and antibonding orbitals. For the retinal model, we employ a minimal CAS(6,6) active space 
of 6 $\pi$ electrons in the orbitals constructed from the $2p_z$ atomic orbitals.   

The CIPSI calculations are performed with Quantum Package~\cite{scemama2019} 
and the determinantal expansions
are constructed to be eigenstates of ${\hat S}^{2}$.  For the cyanine dye where ground and excited
states have different symmetry, we follow two paths to construct the CIPSI expansions: i) we perform
separate expansions for the two states starting from the corresponding CASSCF(6,10) orbitals, and match
the variances of the CI wave functions to obtain a balanced description of the states. We find that
this procedure leads to an automatic match of the second-order perturbation theory (PT2) 
energy contributions as shown in Table S1. ii) We perform the expansion of the two states simultaneously, using
a common set of orbitals (the excited-state CASSCF(6,10) orbitals), and obtain automatically matched
PT2 energy corrections during the expansion~\cite{dash2019}.  For the retinal model where the ground and
excited states have the same symmetry, we have only one set of orbitals for the CIPSI expansions.
In this case, we perform a simultaneous expansion with a selection scheme that matches the CI variances and 
also attempts to balance the PT2 energy contributions of the two states (see SI) \cite{dash2020}.

All total energies are computed on the PBE0/cc-pVQZ ground-state geometry of the cyanine~\cite{boulanger2014} and
retinal molecules.  The DFT geometry optimization of the retinal model  
is performed with the program Gaussian~\cite{Gaussian09}.  
The coupled cluster results are obtained with Psi4~\cite{psi4}. 

\section{Results}
\label{sec:results}

\begin{figure}[bt]
\includegraphics[width=0.45\textwidth]{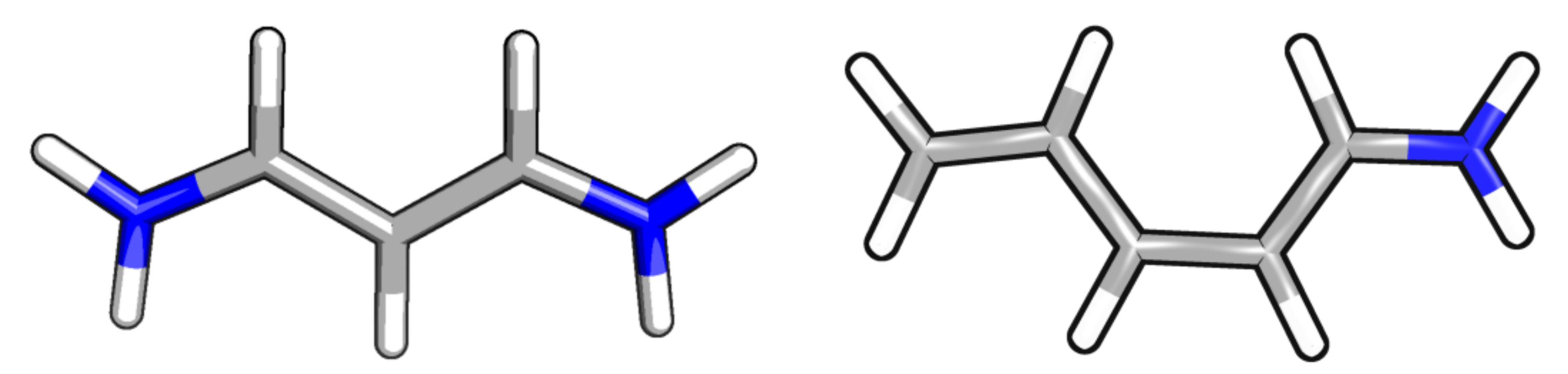}
\caption{Schematic representations of the CN5 (left) and PSB3 (right) molecules. White, gray, and blue
denote hydrogen, carbon, and nitrogen, respectively.}
\label{fig:cn5-psb3}
\end{figure}

We compute the lowest $\pi \rightarrow \pi^{*}$ vertical excitation energy of 
the cyanine dye (C$_3$H$_3$(NH$_2)_2^+$) and the minimal 
model of the retinal protonated Schiff base (C$_5$H$_6$NH$_2^+$) depicted in Fig.~\ref{fig:cn5-psb3}
and denoted as CN5 and PSB3, respectively.  
As already mentioned, while being generally challenging for electronic structure 
methods~\cite{send2011,valsson2012,huix2013,tuna2015,Guennic2015}, these examples are representative of the two
cases of a ground (S0) and an excited (S1) state of different (CN5) and equal (PSB3) symmetry. 
Correspondingly, the energy minimization scheme is state-specific for CN5 and state-average for PSB3, while
variance minimization affords a state-specific optimization
for both molecules, at least in principle. 

\subsection*{Ground and excited states of different symmetry}

\begin{figure}[!htb]
\includegraphics[width=1.0\columnwidth]{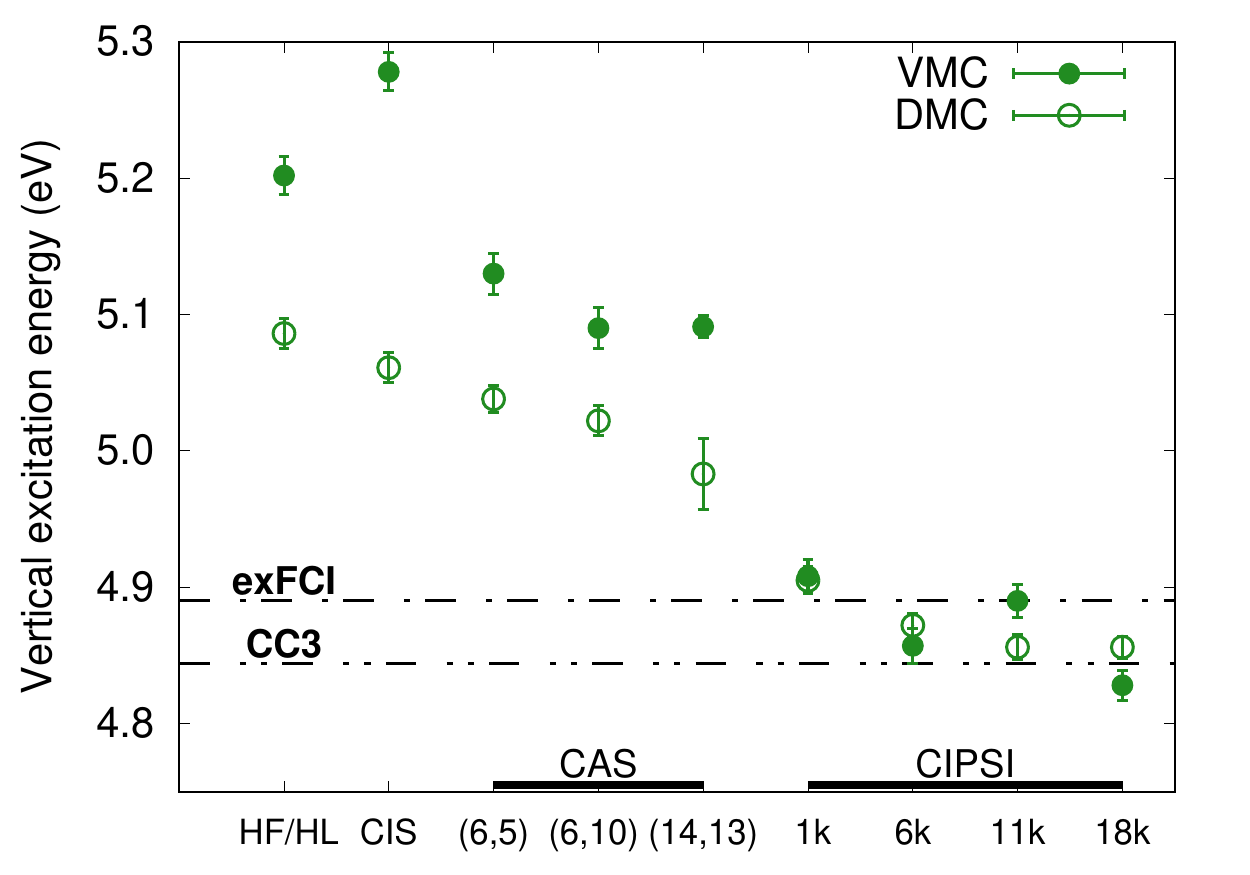}
\caption{
VMC and DMC excitation energies of CN5 calculated with different wave functions optimized in 
energy minimization. The exFCI/aug-cc-pVDZ~\cite{garniron2018} and CC3/aug-cc-pVTZ reference values are
also shown. The approximate total 
number of determinants for the CIPSI-SS wave functions of the ground  and excited states is indicated.}
\label{fig:cn5-energy}
\end{figure}

In Table~\ref{tab:t-cipsi-ene-cn5}, we list the ground- and excited-state energies, and corresponding 
excitation energies of CN5 computed in VMC and DMC with different wave functions optimized by (state-specific) energy minimization.  
The simplest case consists of a single determinant (HF) and a HOMO-LUMO (HL) two-determinant wave function for the ground
and the excited state, respectively. We then consider configuration interaction singles (CIS) expansions, 
CAS expansions with increasing active spaces, and balanced CIPSI expansions with different choices of the 
starting orbitals, namely, independent sets for the two states (CIPSI-SS) or a common set of 
orbitals (CIPSI-B$_1$).  The excitation energies are displayed in Fig.~\ref{fig:cn5-energy}. 

The general trend is a decrease of the excitation energy towards the extrapolated full CI (exFCI) and 
approximate coupled cluster singles, doubles and triples model (CC3) 
reference values for better wave functions.  As an exception, when we move from the HF/HL to CIS 
wave functions, the VMC energies of both states decrease but the corresponding excitation energy becomes 
worse. With increasingly large CAS expansions, both the total and the excitation energies improve but the 
convergence is very slow. For all these wave functions, the DMC excitation energy is lower than the VMC value
and becomes within $0.1$ eV of the reference results for the largest active spaces with about 50,000 and 
70,000 determinants for the ground and the excited state, respectively.  By comparison, the errors of 
TDDFT and CASPT2 can be as large as 0.4 and $-$0.2 eV, respectively~\cite{boulanger2014, send2011}.

\begin{table*}[!htb]
\caption{VMC and DMC total energies (a.u.) and excitation energies ($\Delta$E, eV) of CN5 obtained for 
different wave functions optimizing  all parameters (Jastrow, orbital, 
and CI coefficients) in energy minimization.}
\begin{tabular}{lcccccccccc}
\hline
WF &  \multicolumn{2}{c}{No.\ det} & \multicolumn{2}{c}{No.\ parm} & \multicolumn{2}{c}{E$_{\rm VMC}$} & $\Delta$E$_{\rm VMC}$ & \multicolumn{2}{c}{E$_{\rm DMC}$} & $\Delta$E$_{\rm DMC}$ \\ 
\cline{6-7}
\cline{9-10}
              & S0   & S1     & S0   &  S1  &  S0 & S1 & &  S0 & S1 & \\ 
\hline  
HF/HL        & 1     & 2      & 516  & 529   & -40.8372(4) & -40.6460(3) & 5.202(14) & -40.9378(3) & -40.7509(3) & 5.086(11) \\
HF/CIS       & 1     & 980    & 516  & 4751  & -40.8372(4) & -40.6505(3) & 5.080(14) & -40.9378(3) & -40.7533(3) & 5.020(11) \\          
CIS          & 999   & 980    & 5260 & 4751  & -40.8444(4) & -40.6505(3) & 5.278(14) & -40.9393(3) & -40.7533(3) & 5.061(11) \\ 
CAS(6,5)     & 52    & 48     & 567  & 561   & -40.8468(4) & -40.6583(4) & 5.130(15) & -40.9433(3) & -40.7582(2) & 5.038(10) \\
CAS(6,10)    & 7232  & 7168   & 3134 & 3064  & -40.8498(4) & -40.6628(4) & 5.090(15) & -40.9439(3) & -40.7594(3) & 5.022(11) \\
CAS(14,13)   & 48206 & 72732  & 9480 & 11727 & -40.8583(3) & -40.6713(3) & 5.091(10) & -40.9442(7) & -40.7611(7) & 4.983(26)\\  
CIPSI-SS & 376  & 1094    & 1567 & 2609 & -40.8646(3) & -40.6842(3) & 4.908(12) & -40.9467(3) & -40.7665(3) & 4.905(10) \\    
         & 1344 & 4382    & 2478 & 4531 & -40.8798(3) & -40.7013(3) & 4.857(13) & -40.9502(2) & -40.7711(2) & 4.872(09) \\
         & 2460 & 8782    & 3555 & 6561 & -40.8896(3) & -40.7099(3) & 4.890(12) & -40.9532(2) & -40.7748(2) & 4.856(09) \\
         & 3913 & 14114   & 4842 & 8312 & -40.8941(2) & -40.7167(3) & 4.828(11) & -40.9559(2) & -40.7775(2) & 4.856(08) \\
CIPSI-B$_1$ & 2456 & 6120    & 3971 & 5466 & -40.8847(2) & -40.7053(2) & 4.880(09) & -40.9521(2) & -40.7727(2) & 4.881(09) \\
            & 4829 & 13130   & 5737 & 8021 & -40.8945(3) & -40.7150(3) & 4.889(13) & -40.9560(2) & -40.7766(2) & 4.882(08) \\
\hline
\multicolumn{10}{l}{exFCI/aug-cc-pVDZ~\cite{garniron2018}}  &  4.89  \\
\multicolumn{10}{l}{CC3/aug-cc-pVDZ}                        &  4.851 \\
\multicolumn{10}{l}{CC3/aug-cc-pVTZ}                        &  4.844 \\
\hline
\end{tabular}
\label{tab:t-cipsi-ene-cn5}
\end{table*}

The quality of the results exhibits a further, dramatic improvement with the use of CIPSI expansions. 
The VMC and DMC energies obtained with the smallest CIPSI wave function are lower than the corresponding values
obtained with the largest CAS considered here.  Furthermore, constructing ground- and excited-state 
CIPSI expansions with similar PT2 corrections leads to a balanced description of both states and to
VMC excitation energies which change very little with increasing expansion size, being irregularly scattered 
over a small energy range of 0.08 eV.  Importantly, the DMC excitation energies are compatible 
with the VMC ones and in excellent agreement with the CC3 and exFCI values.
Finally, employing two different sets of orbitals to generate the CIPSI expansions leads to marginal differences, namely,
to DMC excitation energies of 4.856(8) and 4.882(8) eV, which are both bracketed by the reference values.

\begin{figure}[!htb]
\includegraphics[width=1.0\columnwidth]{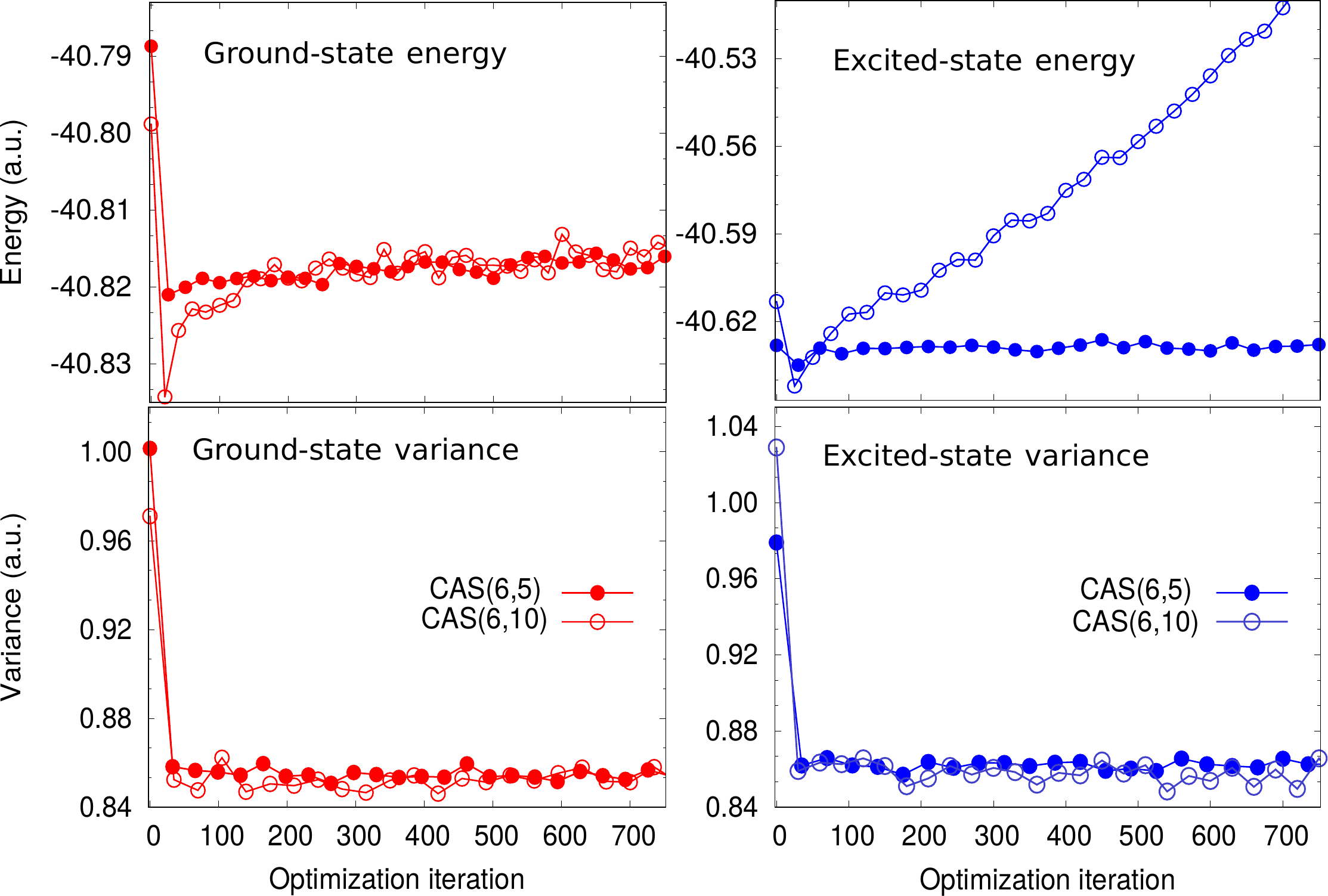}
\caption{Convergence of the VMC energy (top) and variance (bottom) of the ground (left) and excited (right) states of CN5 in
the optimization of the CAS(6,5) and CAS(6,10) wave functions in variance minimization.}
\label{fig:cn5-cas-var}
\end{figure}

Having verified that state-specific energy optimization in combination with accurate wave functions
allows the robust treatment of CN5, we now employ variance minimization with the $\sigma^2_{\omega}$ functional
to optimize the CAS(6,5) and CAS(6,10) wave functions of the ground and excited states. The convergence of the 
corresponding VMC variances and energies is shown in Fig.~\ref{fig:cn5-cas-var}. For the smaller CAS(6,5), we
observe that, while the variance converges rather quickly, the energy appears to do so more slowly and only after undershooting 
to a value which generally depends on the statistical error and initial conditions of the run. As reported
in Table~\ref{tab:t-var-ene-cn5}, the optimal ground- and excited-state energies are higher by about 30 mHartree than the 
corresponding values obtained in energy minimization but the resulting excitation energy is compatible 
within statistical error. 

\begin{figure}[!htb]
\includegraphics[width=1.0\columnwidth]{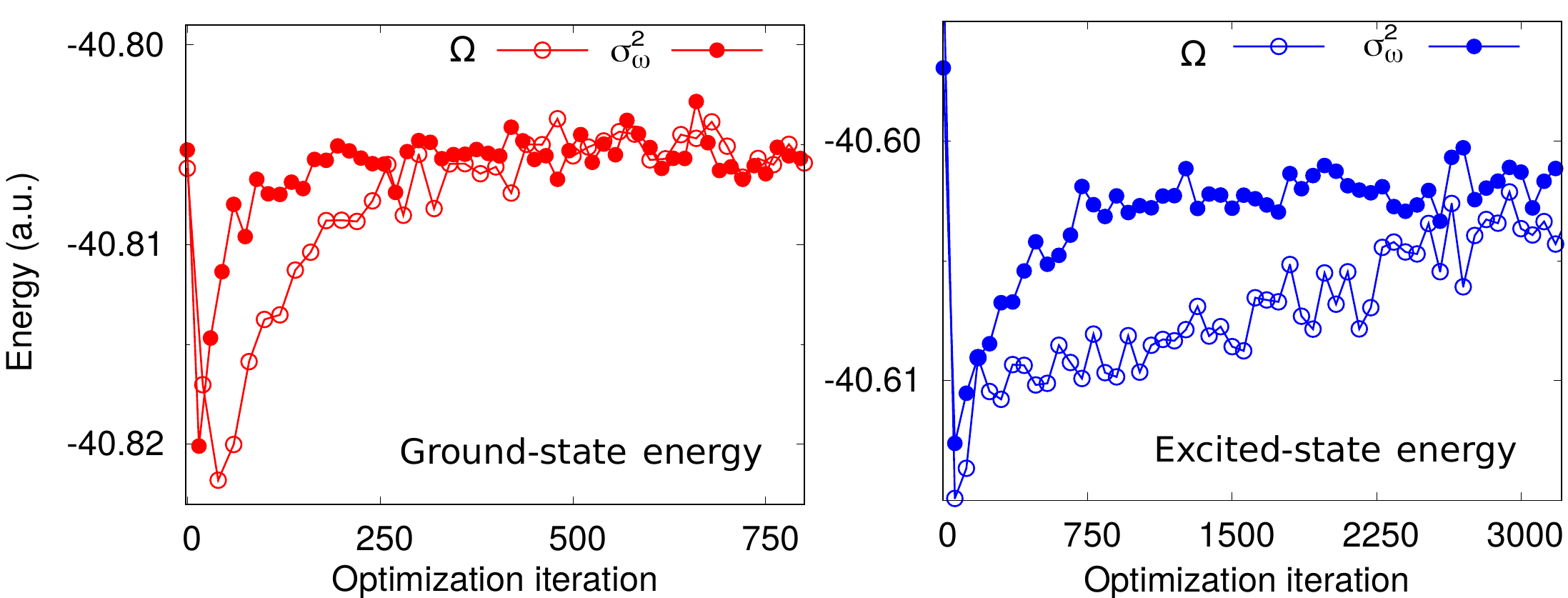}
\caption{Convergence of VMC energy of the ground (left) and excited (right) states of CN5 in the optimization of the HF/HL wave functions 
within variance minimization with the $\sigma^2_\omega$ (our default) and the $\Omega$ functional.}
\label{fig:cn5-rhf-hl}
\end{figure}

If we move to the larger CAS(6,10) determinantal expansion, we find however that, while the variance reaches a 
stable value and the ground-state energy has a similar behavior to the CAS(6,5) case, the energy of the 
excited state grows steadily and it is therefore not possible to estimate the vertical excitation energy 
of the system.  Surprisingly, even in the simplest case of the one-configuration (HF/HL) wave functions, the energy 
of the excited state keeps slowly rising even after 600 iterations as shown in Fig.~\ref{fig:cn5-rhf-hl}, while the ground-state energy 
behaves similarly to the corresponding CAS cases.

\begin{table*}[!htb]
\caption{VMC energies and variances (a.u.) and vertical excitation energies (eV) of CN5 obtained with energy and variance minimization.}
\begin{tabular}{lcccccccccccc}
\hline
          & \multicolumn{5}{c}{Energy min.} && \multicolumn{5}{c}{Variance min.} \\
\cline{2-6}\cline{8-12}
              &  E(S0)        & E(S1)    & $\Delta$E   & $\sigma^2$(S0) & $\sigma^2$(S1)  && E(S0)       &  E(S1)      & $\Delta$E   & $\sigma^2$(S0) & $\sigma^2$(S1) \\
\hline
CAS(6,5)      &  -40.8468(4)  &  -40.6583(4) & 5.13(1)   & 0.928 & 0.940  && -40.8170(5) & -40.6270(5) & 5.17(2) & 0.856 & 0.862 \\
CAS(6,10)     &  -40.8498(4)  &  -40.6628(4) & 5.09(1)   & 0.928 & 0.930  && -40.8163(4) &     --      &   --    & 0.855 &  --  \\
\hline
\end{tabular}
\label{tab:t-var-ene-cn5}
\end{table*}  

We stress that, for most wave functions of Table~\ref{tab:t-cipsi-ene-cn5}, we observe 
a steady growth of the excited-state energy in variance minimization and, therefore, wonder whether
optimizations which appear converged are not simply affected by a much slower drift in the energy, which would
become evident only in longer runs.  Importantly, the apparently unstable behavior is independent of the initial value of $\omega$ 
and the number of steps over which we keep $\omega$ fixed (see Section S6).  The use of a smaller or larger damping 
factors (i.e.\ $\tau =$ 0.04 and 0.2) leads to the same pathological growth of the excited-state energy, characterized 
by the same slope as a function of time as shown in Fig.~S4. 
Moreover, we recover the same behavior also when using a gradient-only-based optimizer (see Fig.~S7).  
Finally, minimizing the $\Omega$ functional instead of $\sigma^2_{\omega}$ yields an excited-state energy 
which ultimately rises with iterations as shown for the excited-state HL wave function 
in Fig.~\ref{fig:cn5-rhf-hl}.

\subsection*{Ground and excited states of the same symmetry}

For PSB3, we optimize the wave functions in energy minimization in a state-average fashion
and report the resulting VMC and DMC total energies and vertical excitation energies 
in Table~\ref{tab:tbpsb3}.
As in the CN5 case, CIPSI wave functions are superior to CAS expansions of similar size and, with only
about 400 determinants, the use of CIPSI yields not only lower total energies but also a VMC vertical excitation 
energy in good agreement with the CC3 reference, largely correcting the error of 0.25 eV obtained with the CAS(6,6) wave
function.  For all CIPSI expansions, the DMC excitation energies are always quite close
to the correspondent VMC results and, for the larger expansions, within 0.05 eV of the CC3 value.

\begin{table*}[!htb]
\caption{VMC and DMC total energies (a.u.) and vertical excitation energies (eV) of PSB3 obtained with 
different wave functions optimized in energy minimization.}
\begin{tabular}{lccccccccc}
\hline
WF &  No.\ det &   No.\ param  & \multicolumn{3}{c}{VMC} && \multicolumn{3}{c}{DMC} \\ 
\cline{4-6}\cline{8-10}
         &      &      & E(S0)       & E(S1)       &$\Delta$E && E(S0)       &  E(S1)      & $\Delta$E \\ 
\hline
CAS(6,6) & 400  & 1645  & -42.8091(2) & -42.6471(2) & 4.409(9) && -42.9118(2) & -42.7541(2) & 4.293(6) \\
CIPSI    & 422  & 4011  & -42.8174(2) & -42.6623(2) & 4.221(9) && -42.9133(2) & -42.7578(2) & 4.233(6) \\
         & 1158 & 5968  & -42.8297(2) & -42.6735(2) & 4.252(9) && -42.9160(2) & -42.7609(2) & 4.221(6)\\
         & 2579 & 8106  & -42.8357(2) & -42.6796(2) & 4.247(9) && -42.9169(2) & -42.7621(2) & 4.214(6) \\
\hline
\multicolumn{9}{l}{CC3/aug-cc-pVDZ}   & 4.19 \\
\multicolumn{9}{l}{CC3/aug-cc-pVTZ}   & 4.16 \\
\hline
\end{tabular}
\label{tab:tbpsb3}
\end{table*}

\begin{figure}[!htb]
\includegraphics[width=1.0\columnwidth]{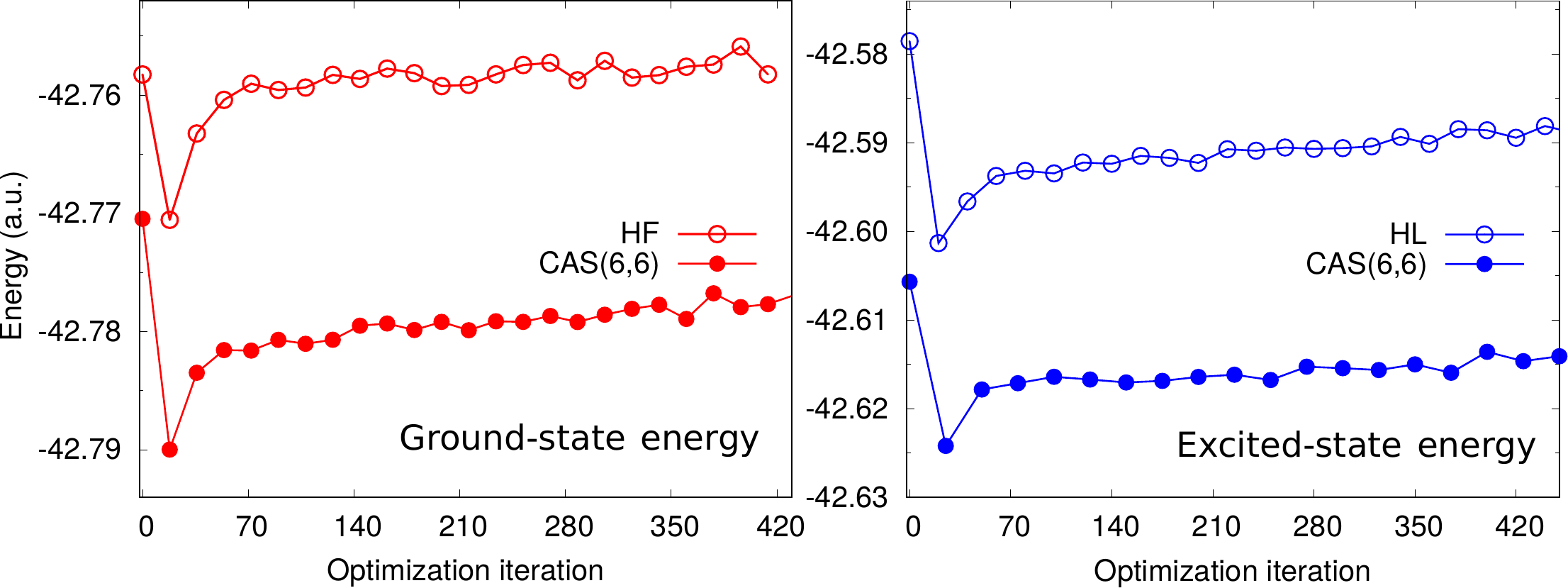}
\caption{Convergence of the VMC energy of the ground (red) and excited (blue) states of PSB3 in the optimization 
of the RHF/HL and CAS(6,6) wave functions within variance minimization.}
\label{fig:psb3-conv-lvmq}
\end{figure}

When we perform state-specific variance minimization, we encounter great difficulties in the convergence
of the energies as we show for the HF/HL and CAS(6,6) wave functions in Fig.~\ref{fig:psb3-conv-lvmq}. Differently
from CN5, we find in general that not only the energy of the excited state but also that of the ground state 
grows steadily with iteration number.  We only obtain a seemingly stable energy for the HF wave function.

\section{Discussion and conclusions}
\label{sec:disc}

While our results confirm the high accuracy reachable in QMC with energy minimization, they evidence 
severe problems in variance minimization which, in most cases, preclude the estimation of the excitation 
energy.  To gain a better understanding of the troublesome behavior of the energy during variance minimization, we 
further investigate the simple case of the HL wave function of CN5 (Fig.~\ref{fig:cn5-rhf-hl}) and find that the energy 
of the state drifts to higher values during variance minimization also when one optimizes only the LUMO orbital. 
Therefore, since optimization of an orbital can be achieved by mixing it with the unoccupied ones 
of the same symmetry, we can recast the LUMO optimization into the linear variation of the CI coefficients 
of the single excitations out of the LUMO orbital, which amount to only twelve additional CSFs in our basis set.
With such a small expansion, we can then diagonalize the Hamiltonian in the basis of the CSFs times the Jastrow factor 
to estimate its thirteen eigenvalues and eigenvectors, and work directly in the basis of the eigenstates to assess
the behavior of variance minimization when starting from the states which are optimal for energy minimization. 
  
\begin{figure}[!htb]
\includegraphics[width=1.0\columnwidth]{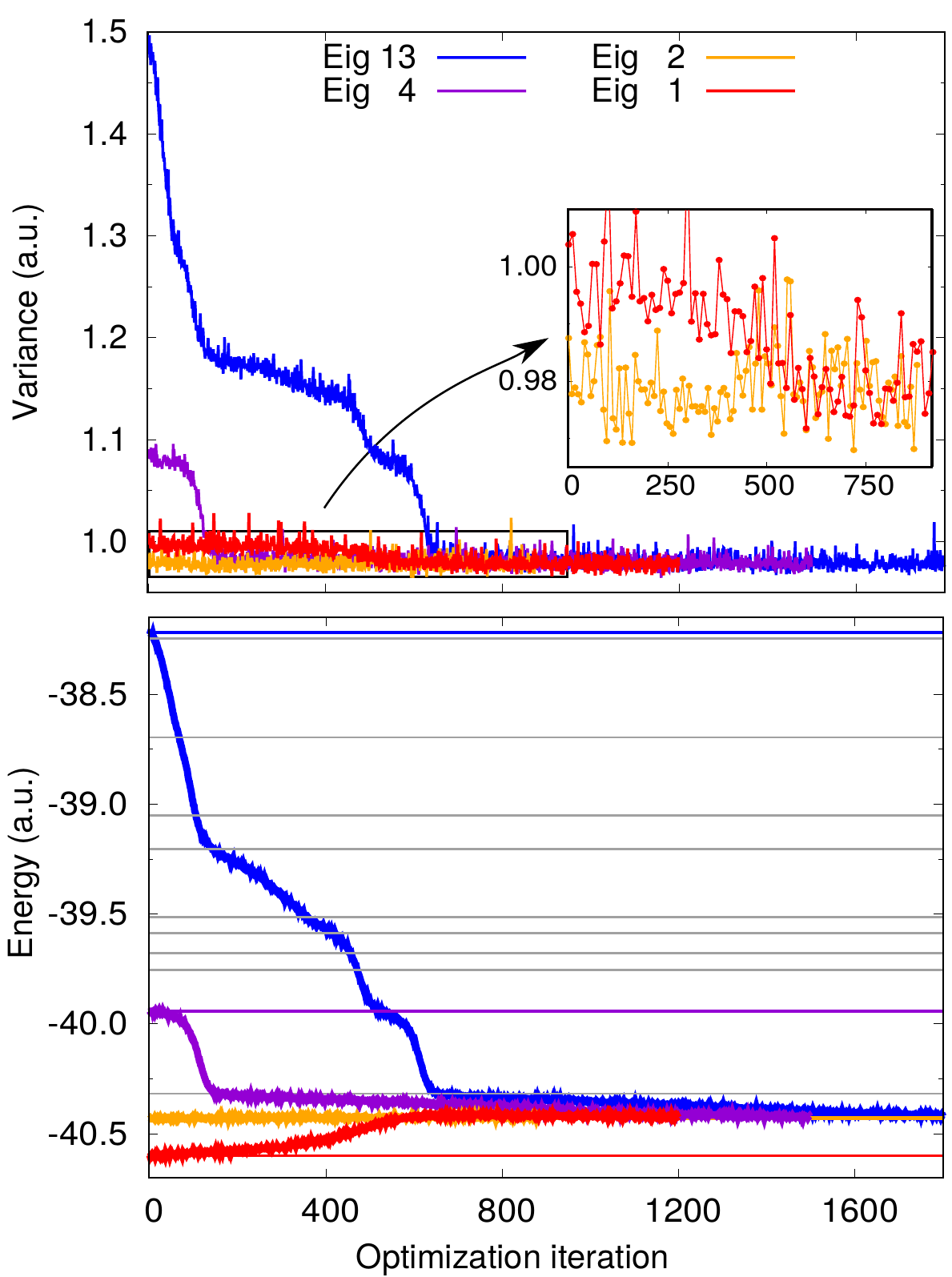}
\caption{Convergence of the VMC variance (top) and energy (bottom) of CN5 in the CI optimization of a small expansion 
(see text) with variance minimization. The horizontal lines in the energy plot correspond to the eigenvalues in this 
reduced space, and the colored ones are the eigenstates used as starting point in four optimization runs. The damping
factor used in the Newton method is $\tau=0.2$.}
\label{fig:lumo-opt}
\end{figure}

In Fig.~\ref{fig:lumo-opt}, we show the evolution of the VMC variance and energy for four variance 
minimization runs in which we start from different eigenvectors, taking the corresponding eigenvalues as 
initial target energies $\omega$.  In particular, we consider the lowest state in B$_{1}$ symmetry as well 
as the second, fourth, and thirteenth (corresponding to the highest energy) states.  
We note that, since our states are not exact eigenstates of the full Hamiltonian, the corresponding variances 
of the local energy are non zero and are spread over about 0.5 a.u.\ with the lowest value in correspondence of
the second state. In principle, one would expect to find a feature of the variance landscape
--ideally a local minimum-- near each of the approximate eigenstates since the functionals 
$\sigma^2_\omega$ or $\Omega$ are designed to select a particular state through the initial 
value of $\omega$, and minimize the variance of {\em that} state. Here, the selection of the 
state is further facilitated starting each run precisely from the chosen
eigenstate, and variance minimization should perform minor adjustments
of the initial parameters from their optimal values for the energy.

\begin{figure}[!htb]
\includegraphics[width=1\columnwidth]{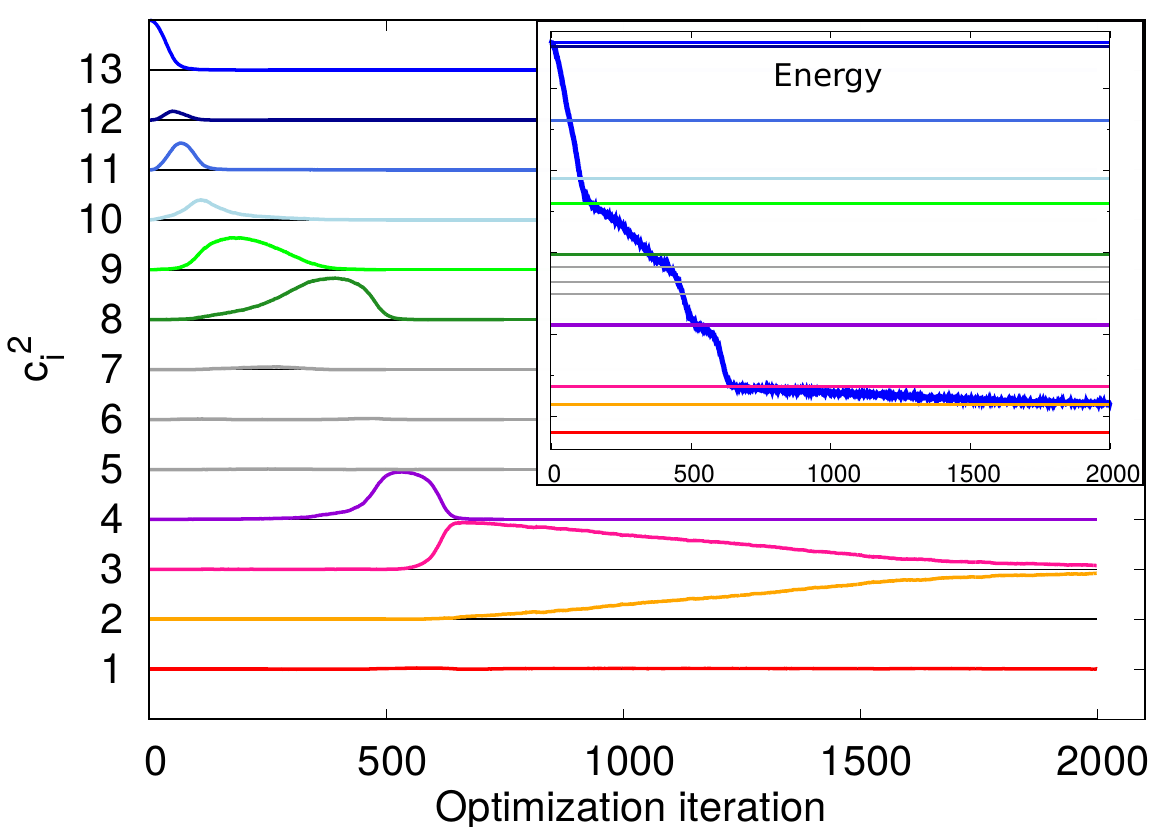}
\caption{
Evolution of the square of the CI coefficients 
$c_i^2$ (offset by $i$ for clarity) of the small expansion 
of CN5 during variance minimization, for the run starting from 
the 13$^{\text{th}}$ eigenvector; in the inset, the
evolution of the energy is replicated to emphasize flat regions 
in the energy landscape close
to an eigenstate (i.e.\ when the corresponding $c_i\sim 1$).
}
\label{fig:lumo-opt-ci2}
\end{figure}

The behavior illustrated in Figs.~\ref{fig:lumo-opt}
is totally different, with all optimization runs leaking down to 
successive lower-variance states and eventually converging to the 
absolute minimum corresponding to the second eigenstate.
The staircase shape of the variance evolution points to the presence of flat regions of the 
variance landscape close to the eigenstates, from which the optimization can eventually escape.
This is further corroborated if we follow the evolution of the CI coefficients
as shown starting from the highest-energy state in Fig.~\ref{fig:lumo-opt-ci2}:
the initial coefficient quickly decreases to zero and other eigenstates become
populated until convergence on the second state. In proximity of some eigenstates, the variance displays
a more pronounced plateau, where the system spends enough time to acquire the full
character of that particular state.
It is also interesting to note that the states are populated sequentially with 
the order determined by decreasing energies.
We stress that we observe a similar behavior of the variance also when using
the $\Omega$ functional starting from the same set of approximate eigenstates  (see Fig.~S8).

\begin{figure}[!htb]
\includegraphics[width=1.\columnwidth]{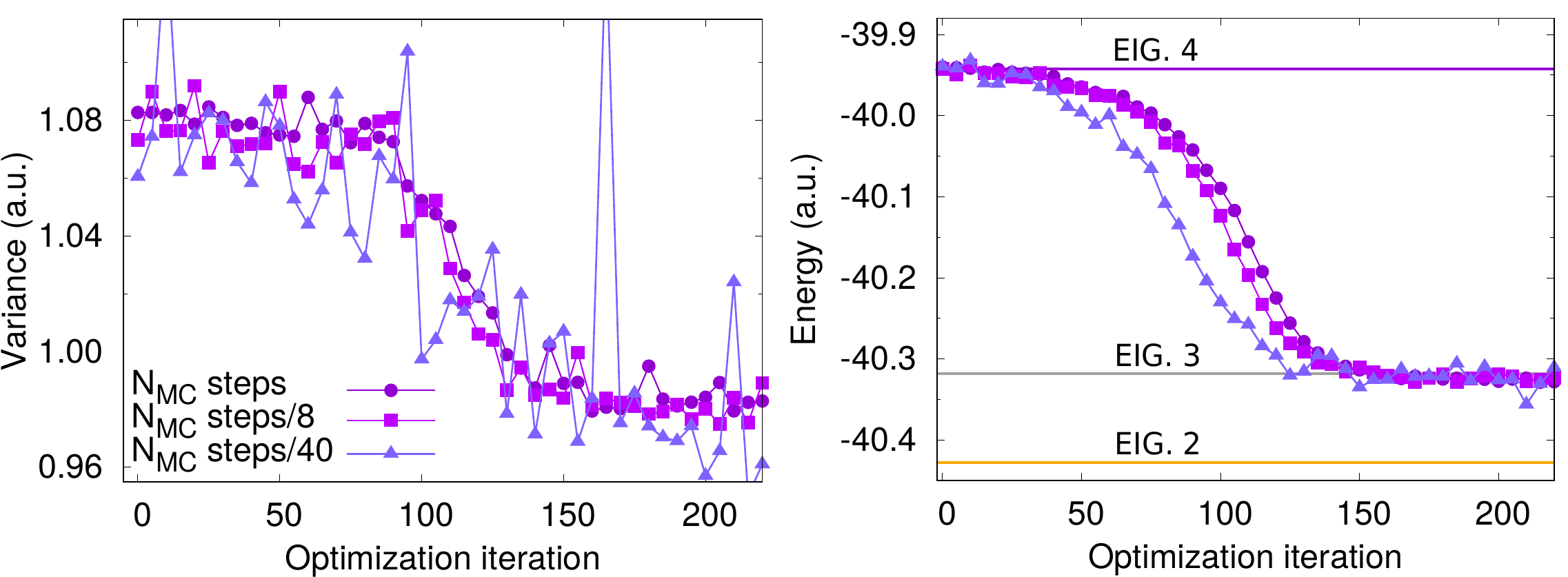}
\caption{Convergence of the variance (left) and energy (right) for different lengths of the Monte Carlo runs used to 
compute the gradient and Hessian matrix during the optimization, starting from the $4^{\text{th}}$ eigenvector.
$N_\text{{MC}}$ is the number of Monte Carlo steps used in Fig.~\ref{fig:lumo-opt}.
}
\label{fig:lumo-opt-stat}
\end{figure}

In Fig.~\ref{fig:lumo-opt-stat}, we 
investigate the impact of the statistical error on the loss 
of the selected state.
In particular, we focus on the evolution of the variance and 
the energy starting from the $4^{\text{th}}$ eigenvector 
for different lengths of the VMC runs used to compute the 
gradient and Hessian matrix. The shortest run (larger 
statistical error) looses the target state in a slightly smaller 
number of steps. However the intermediate and the longest run
give very similar results, suggesting that even longer VMC runs would
not stabilize the target state. 

\begin{figure}[!htb]
\includegraphics[width=1.0\columnwidth]{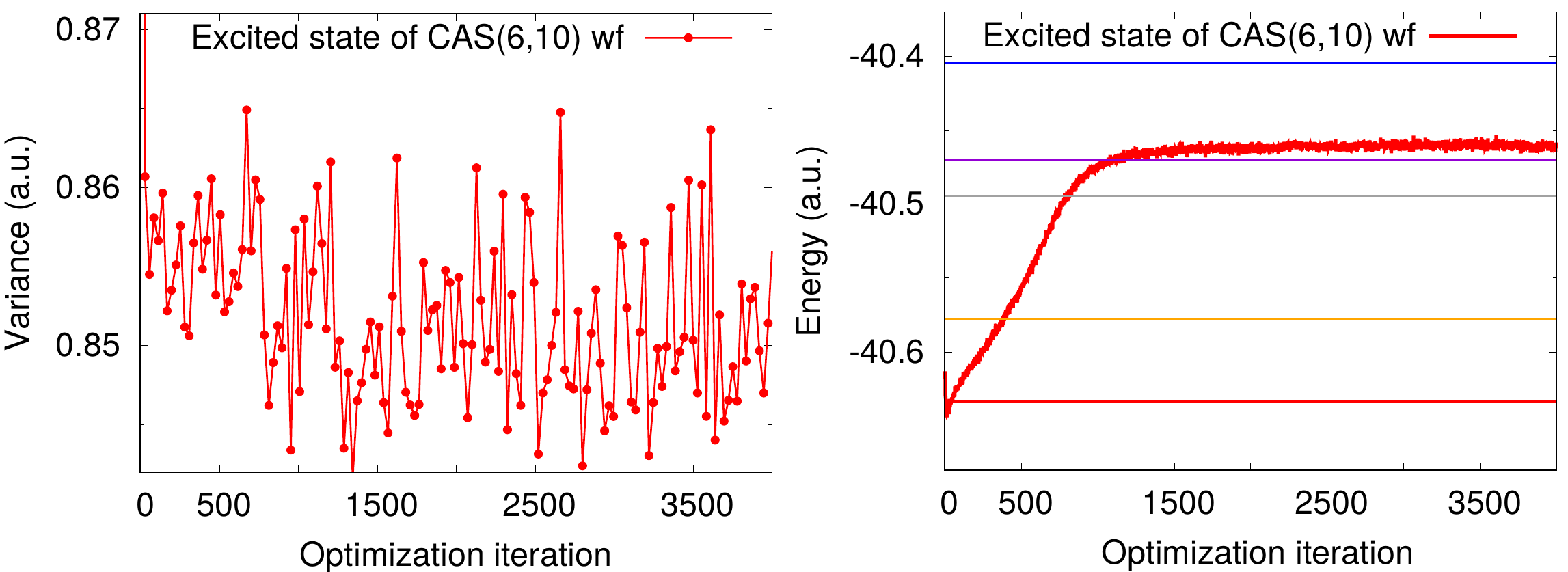}
\caption{Variance (left) and energy (right) convergence for the optimization of the excited state of
the CAS(6,10) wave function. The horizontal lines in the energy plot correspond to the firsts eigenvalue
roots obtained with the Davidson optimization.}
\label{fig:cas610-eigen}
\end{figure}

Finally, to verify that the understanding gained here also applies to more complicated wave functions, we revisit 
the very problematic optimization of the excited-state CAS(6,10) wave function (Fig.~\ref{fig:cn5-cas-var}) and 
perform a much longer calculation, finding that the energy eventually converges as shown in Fig.~\ref{fig:cas610-eigen}. 
For the final set of Jastrow and orbital parameters, we determine the eigenvalues in the linear space of the determinants 
times the Jastrow factor and recover a similar behavior to what observed in the simple example:
the minimization of $\sigma_{\omega}^2$ brings the system approximately to an eigenstate with a lower variance, 
which is in this case the 4$^{\text{th}}$ one.

In summary, we have shown that the combination of energy minimization with an appropriate choice of the
ground- and excited-state wave functions via a balanced CIPSI procedure leads to excitation energies that are in excellent 
agreement already at the VMC level with the reference values. In particular, we obtained a 
robust convergence of the total ground- and excited-state energies, and a very accurate excitation 
energy not only in the easier state-specific case of CN5 but also when employing energy minimization in a 
state-average fashion for PSB3.  On the other hand, we encountered severe problems when employing variance
minimization since, over sufficiently long optimization runs, one may loose the state of interest in favor 
of a state with lower variance, as we clearly demonstrated with a simple but realistic example. 
Even though, theoretically, the functionals $\sigma_\omega^2$ and $\Omega$ have a built-in possibility 
to target the energy of a specific state, in practice, this is generally not sufficient to maintain the
parameters close to the desired local minimum of the variance. 
Therefore, these considerations lead to the conclusion that, with the present functionals and no
{\it a priori} knowledge of the parameter landscape of the variance for the system of interest, energy minimization 
is a safer and more stable procedure. 

\section*{Acknowledgment}

A.C.\ is supported by the ``Computational Science for Energy Research and Netherlands eScience Center joint program'' 
(project CSER.JCER.022) of the Netherlands Organisation for Scientific Research (NWO).
This work was carried out on the Dutch national supercomputer Cartesius with the support of SURF Cooperative.

\section*{Content of SI}

Derivation and discussion of the expressions of the gradient and approximate Hessian of the variance; 
CIPSI energies for various expansions; basis-set dependence of the VMC and DMC excitation energies; 
DMC excitation energy versus time step; dependence of variance minimization on the choice of $\omega$, number of 
steps with $\omega$ fixed, damping factor in the Newton method, and statistical error;
optimizations with a gradient-based optimizer and with the $\Omega$ functional.

\bibliography{paper_enevarQMC_Cuzzocrea}

\begin{thebibliography}{48}%
\makeatletter
\providecommand \@ifxundefined [1]{%
 \@ifx{#1\undefined}
}%
\providecommand \@ifnum [1]{%
 \ifnum #1\expandafter \@firstoftwo
 \else \expandafter \@secondoftwo
 \fi
}%
\providecommand \@ifx [1]{%
 \ifx #1\expandafter \@firstoftwo
 \else \expandafter \@secondoftwo
 \fi
}%
\providecommand \natexlab [1]{#1}%
\providecommand \enquote  [1]{``#1''}%
\providecommand \bibnamefont  [1]{#1}%
\providecommand \bibfnamefont [1]{#1}%
\providecommand \citenamefont [1]{#1}%
\providecommand \href@noop [0]{\@secondoftwo}%
\providecommand \href [0]{\begingroup \@sanitize@url \@href}%
\providecommand \@href[1]{\@@startlink{#1}\@@href}%
\providecommand \@@href[1]{\endgroup#1\@@endlink}%
\providecommand \@sanitize@url [0]{\catcode `\\12\catcode `\$12\catcode
  `\&12\catcode `\#12\catcode `\^12\catcode `\_12\catcode `\%12\relax}%
\providecommand \@@startlink[1]{}%
\providecommand \@@endlink[0]{}%
\providecommand \url  [0]{\begingroup\@sanitize@url \@url }%
\providecommand \@url [1]{\endgroup\@href {#1}{\urlprefix }}%
\providecommand \urlprefix  [0]{URL }%
\providecommand \Eprint [0]{\href }%
\providecommand \doibase [0]{http://dx.doi.org/}%
\providecommand \selectlanguage [0]{\@gobble}%
\providecommand \bibinfo  [0]{\@secondoftwo}%
\providecommand \bibfield  [0]{\@secondoftwo}%
\providecommand \translation [1]{[#1]}%
\providecommand \BibitemOpen [0]{}%
\providecommand \bibitemStop [0]{}%
\providecommand \bibitemNoStop [0]{.\EOS\space}%
\providecommand \EOS [0]{\spacefactor3000\relax}%
\providecommand \BibitemShut  [1]{\csname bibitem#1\endcsname}%
\let\auto@bib@innerbib\@empty
\bibitem [{\citenamefont {Filippi}\ \emph {et~al.}(2009)\citenamefont
  {Filippi}, \citenamefont {Zaccheddu},\ and\ \citenamefont
  {Buda}}]{filippi2009}%
  \BibitemOpen
  \bibfield  {author} {\bibinfo {author} {\bibfnamefont {C.}~\bibnamefont
  {Filippi}}, \bibinfo {author} {\bibfnamefont {M.}~\bibnamefont {Zaccheddu}},
  \ and\ \bibinfo {author} {\bibfnamefont {F.}~\bibnamefont {Buda}},\
  }\href@noop {} {\bibfield  {journal} {\bibinfo  {journal} {J. Chem. Theory
  Comput.}\ }\textbf {\bibinfo {volume} {5}},\ \bibinfo {pages} {2074}
  (\bibinfo {year} {2009})}\BibitemShut {NoStop}%
\bibitem [{\citenamefont {Zimmerman}\ \emph {et~al.}(2009)\citenamefont
  {Zimmerman}, \citenamefont {Toulouse}, \citenamefont {Zhang}, \citenamefont
  {Musgrave},\ and\ \citenamefont {Umrigar}}]{zimmerman2009}%
  \BibitemOpen
  \bibfield  {author} {\bibinfo {author} {\bibfnamefont {P.~M.}\ \bibnamefont
  {Zimmerman}}, \bibinfo {author} {\bibfnamefont {J.}~\bibnamefont {Toulouse}},
  \bibinfo {author} {\bibfnamefont {Z.}~\bibnamefont {Zhang}}, \bibinfo
  {author} {\bibfnamefont {C.~B.}\ \bibnamefont {Musgrave}}, \ and\ \bibinfo
  {author} {\bibfnamefont {C.}~\bibnamefont {Umrigar}},\ }\href@noop {}
  {\bibfield  {journal} {\bibinfo  {journal} {J. Chem. Phys.}\ }\textbf
  {\bibinfo {volume} {131}},\ \bibinfo {pages} {124103} (\bibinfo {year}
  {2009})}\BibitemShut {NoStop}%
\bibitem [{\citenamefont {Valsson}\ and\ \citenamefont
  {Filippi}(2010)}]{valsson2010}%
  \BibitemOpen
  \bibfield  {author} {\bibinfo {author} {\bibfnamefont {O.}~\bibnamefont
  {Valsson}}\ and\ \bibinfo {author} {\bibfnamefont {C.}~\bibnamefont
  {Filippi}},\ }\href@noop {} {\bibfield  {journal} {\bibinfo  {journal} {J.
  Chem. Theory Comput.}\ }\textbf {\bibinfo {volume} {6}},\ \bibinfo {pages}
  {1275} (\bibinfo {year} {2010})}\BibitemShut {NoStop}%
\bibitem [{\citenamefont {Send}\ \emph {et~al.}(2011)\citenamefont {Send},
  \citenamefont {Valsson},\ and\ \citenamefont {Filippi}}]{send2011}%
  \BibitemOpen
  \bibfield  {author} {\bibinfo {author} {\bibfnamefont {R.}~\bibnamefont
  {Send}}, \bibinfo {author} {\bibfnamefont {O.}~\bibnamefont {Valsson}}, \
  and\ \bibinfo {author} {\bibfnamefont {C.}~\bibnamefont {Filippi}},\ }\href
  {\doibase 10.1021/ct1006295} {\bibfield  {journal} {\bibinfo  {journal}
  {Journal of Chemical Theory and Computation}\ }\textbf {\bibinfo {volume}
  {7}},\ \bibinfo {pages} {444} (\bibinfo {year} {2011})}\BibitemShut {NoStop}%
\bibitem [{\citenamefont {Valsson}\ \emph {et~al.}(2013)\citenamefont
  {Valsson}, \citenamefont {Campomanes}, \citenamefont {Tavernelli},
  \citenamefont {Rothlisberger},\ and\ \citenamefont {Filippi}}]{valsson2013}%
  \BibitemOpen
  \bibfield  {author} {\bibinfo {author} {\bibfnamefont {O.}~\bibnamefont
  {Valsson}}, \bibinfo {author} {\bibfnamefont {P.}~\bibnamefont {Campomanes}},
  \bibinfo {author} {\bibfnamefont {I.}~\bibnamefont {Tavernelli}}, \bibinfo
  {author} {\bibfnamefont {U.}~\bibnamefont {Rothlisberger}}, \ and\ \bibinfo
  {author} {\bibfnamefont {C.}~\bibnamefont {Filippi}},\ }\href@noop {}
  {\bibfield  {journal} {\bibinfo  {journal} {J. Chem. Theory Comput.}\
  }\textbf {\bibinfo {volume} {9}},\ \bibinfo {pages} {2441} (\bibinfo {year}
  {2013})}\BibitemShut {NoStop}%
\bibitem [{\citenamefont {Guareschi}\ \emph {et~al.}(2016)\citenamefont
  {Guareschi}, \citenamefont {Zulfikri}, \citenamefont {Daday}, \citenamefont
  {Floris}, \citenamefont {Amovilli}, \citenamefont {Mennucci},\ and\
  \citenamefont {Filippi}}]{guareschi2016}%
  \BibitemOpen
  \bibfield  {author} {\bibinfo {author} {\bibfnamefont {R.}~\bibnamefont
  {Guareschi}}, \bibinfo {author} {\bibfnamefont {H.}~\bibnamefont {Zulfikri}},
  \bibinfo {author} {\bibfnamefont {C.}~\bibnamefont {Daday}}, \bibinfo
  {author} {\bibfnamefont {F.~M.}\ \bibnamefont {Floris}}, \bibinfo {author}
  {\bibfnamefont {C.}~\bibnamefont {Amovilli}}, \bibinfo {author}
  {\bibfnamefont {B.}~\bibnamefont {Mennucci}}, \ and\ \bibinfo {author}
  {\bibfnamefont {C.}~\bibnamefont {Filippi}},\ }\href@noop {} {\bibfield
  {journal} {\bibinfo  {journal} {Journal of Chemical Theory and Computation}\
  }\textbf {\bibinfo {volume} {12}},\ \bibinfo {pages} {1674} (\bibinfo {year}
  {2016})}\BibitemShut {NoStop}%
\bibitem [{\citenamefont {Hunt}\ \emph {et~al.}(2018)\citenamefont {Hunt},
  \citenamefont {Szyniszewski}, \citenamefont {Prayogo}, \citenamefont
  {Maezono},\ and\ \citenamefont {Drummond}}]{hunt2018}%
  \BibitemOpen
  \bibfield  {author} {\bibinfo {author} {\bibfnamefont {R.~J.}\ \bibnamefont
  {Hunt}}, \bibinfo {author} {\bibfnamefont {M.}~\bibnamefont {Szyniszewski}},
  \bibinfo {author} {\bibfnamefont {G.~I.}\ \bibnamefont {Prayogo}}, \bibinfo
  {author} {\bibfnamefont {R.}~\bibnamefont {Maezono}}, \ and\ \bibinfo
  {author} {\bibfnamefont {N.~D.}\ \bibnamefont {Drummond}},\ }\href@noop {}
  {\bibfield  {journal} {\bibinfo  {journal} {Phys. Rev. B}\ }\textbf {\bibinfo
  {volume} {98}},\ \bibinfo {pages} {075122} (\bibinfo {year}
  {2018})}\BibitemShut {NoStop}%
\bibitem [{\citenamefont {Blunt}\ and\ \citenamefont
  {Neuscamman}(2019)}]{blunt2019}%
  \BibitemOpen
  \bibfield  {author} {\bibinfo {author} {\bibfnamefont {N.~S.}\ \bibnamefont
  {Blunt}}\ and\ \bibinfo {author} {\bibfnamefont {E.}~\bibnamefont
  {Neuscamman}},\ }\href {\doibase 10.1021/acs.jctc.8b00879} {\bibfield
  {journal} {\bibinfo  {journal} {J. Chem. Theory Comput.}\ }\textbf {\bibinfo
  {volume} {15}},\ \bibinfo {pages} {178} (\bibinfo {year} {2019})}\BibitemShut
  {NoStop}%
\bibitem [{\citenamefont {Dash}\ \emph {et~al.}(2019)\citenamefont {Dash},
  \citenamefont {Feldt}, \citenamefont {Moroni}, \citenamefont {Scemama},\ and\
  \citenamefont {Filippi}}]{dash2019}%
  \BibitemOpen
  \bibfield  {author} {\bibinfo {author} {\bibfnamefont {M.}~\bibnamefont
  {Dash}}, \bibinfo {author} {\bibfnamefont {J.}~\bibnamefont {Feldt}},
  \bibinfo {author} {\bibfnamefont {S.}~\bibnamefont {Moroni}}, \bibinfo
  {author} {\bibfnamefont {A.}~\bibnamefont {Scemama}}, \ and\ \bibinfo
  {author} {\bibfnamefont {C.}~\bibnamefont {Filippi}},\ }\href {\doibase
  10.1021/acs.jctc.9b00476} {\bibfield  {journal} {\bibinfo  {journal} {J.
  Chem. Theory Comput.}\ }\textbf {\bibinfo {volume} {15}},\ \bibinfo {pages}
  {4896} (\bibinfo {year} {2019})}\BibitemShut {NoStop}%
\bibitem [{\citenamefont {Foulkes}\ \emph {et~al.}(2001)\citenamefont
  {Foulkes}, \citenamefont {Mitas}, \citenamefont {Needs},\ and\ \citenamefont
  {Rajagopal}}]{Foulkes2001}%
  \BibitemOpen
  \bibfield  {author} {\bibinfo {author} {\bibfnamefont {W.~M.~C.}\
  \bibnamefont {Foulkes}}, \bibinfo {author} {\bibfnamefont {L.}~\bibnamefont
  {Mitas}}, \bibinfo {author} {\bibfnamefont {R.~J.}\ \bibnamefont {Needs}}, \
  and\ \bibinfo {author} {\bibfnamefont {G.}~\bibnamefont {Rajagopal}},\
  }\href@noop {} {\bibfield  {journal} {\bibinfo  {journal} {Reviews of Modern
  Physics}\ }\textbf {\bibinfo {volume} {73}},\ \bibinfo {pages} {33} (\bibinfo
  {year} {2001})}\BibitemShut {NoStop}%
\bibitem [{\citenamefont {L\"uchow}(2011)}]{luchow_quantum_2011}%
  \BibitemOpen
  \bibfield  {author} {\bibinfo {author} {\bibfnamefont {A.}~\bibnamefont
  {L\"uchow}},\ }\href@noop {} {\bibfield  {journal} {\bibinfo  {journal}
  {Wiley Interdisciplinary Reviews: Computational Molecular Science}\ }\textbf
  {\bibinfo {volume} {1}},\ \bibinfo {pages} {388} (\bibinfo {year}
  {2011})}\BibitemShut {NoStop}%
\bibitem [{\citenamefont {Austin}\ \emph {et~al.}(2012)\citenamefont {Austin},
  \citenamefont {Zubarev},\ and\ \citenamefont {Lester}}]{austin_quantum_2012}%
  \BibitemOpen
  \bibfield  {author} {\bibinfo {author} {\bibfnamefont {B.~M.}\ \bibnamefont
  {Austin}}, \bibinfo {author} {\bibfnamefont {D.~Y.}\ \bibnamefont {Zubarev}},
  \ and\ \bibinfo {author} {\bibfnamefont {W.~A.}\ \bibnamefont {Lester}},\
  }\href@noop {} {\bibfield  {journal} {\bibinfo  {journal} {Chemical Reviews}\
  }\textbf {\bibinfo {volume} {112}},\ \bibinfo {pages} {263} (\bibinfo {year}
  {2012})}\BibitemShut {NoStop}%
\bibitem [{\citenamefont {Sorella}\ and\ \citenamefont
  {Capriotti}(2010)}]{sorella2010}%
  \BibitemOpen
  \bibfield  {author} {\bibinfo {author} {\bibfnamefont {S.}~\bibnamefont
  {Sorella}}\ and\ \bibinfo {author} {\bibfnamefont {L.}~\bibnamefont
  {Capriotti}},\ }\href {\doibase 10.1063/1.3516208} {\bibfield  {journal}
  {\bibinfo  {journal} {The Journal of Chemical Physics}\ }\textbf {\bibinfo
  {volume} {133}},\ \bibinfo {pages} {234111} (\bibinfo {year}
  {2010})}\BibitemShut {NoStop}%
\bibitem [{\citenamefont {Neuscamman}\ \emph {et~al.}(2012)\citenamefont
  {Neuscamman}, \citenamefont {Umrigar},\ and\ \citenamefont
  {Chan}}]{neuscamman2012}%
  \BibitemOpen
  \bibfield  {author} {\bibinfo {author} {\bibfnamefont {E.}~\bibnamefont
  {Neuscamman}}, \bibinfo {author} {\bibfnamefont {C.~J.}\ \bibnamefont
  {Umrigar}}, \ and\ \bibinfo {author} {\bibfnamefont {G.~K.-L.}\ \bibnamefont
  {Chan}},\ }\href@noop {} {\bibfield  {journal} {\bibinfo  {journal} {Phys.
  Rev. B}\ }\textbf {\bibinfo {volume} {85}},\ \bibinfo {pages} {045103}
  (\bibinfo {year} {2012})}\BibitemShut {NoStop}%
\bibitem [{\citenamefont {Filippi}\ \emph {et~al.}(2016)\citenamefont
  {Filippi}, \citenamefont {Assaraf},\ and\ \citenamefont
  {Moroni}}]{filippi2016}%
  \BibitemOpen
  \bibfield  {author} {\bibinfo {author} {\bibfnamefont {C.}~\bibnamefont
  {Filippi}}, \bibinfo {author} {\bibfnamefont {R.}~\bibnamefont {Assaraf}}, \
  and\ \bibinfo {author} {\bibfnamefont {S.}~\bibnamefont {Moroni}},\
  }\href@noop {} {\bibfield  {journal} {\bibinfo  {journal} {J. Chem. Phys.}\
  }\textbf {\bibinfo {volume} {144}},\ \bibinfo {pages} {194105} (\bibinfo
  {year} {2016})}\BibitemShut {NoStop}%
\bibitem [{\citenamefont {Assaraf}\ \emph {et~al.}(2017)\citenamefont
  {Assaraf}, \citenamefont {Moroni},\ and\ \citenamefont
  {Filippi}}]{assaraf2017}%
  \BibitemOpen
  \bibfield  {author} {\bibinfo {author} {\bibfnamefont {R.}~\bibnamefont
  {Assaraf}}, \bibinfo {author} {\bibfnamefont {S.}~\bibnamefont {Moroni}}, \
  and\ \bibinfo {author} {\bibfnamefont {C.}~\bibnamefont {Filippi}},\
  }\href@noop {} {\bibfield  {journal} {\bibinfo  {journal} {J. Chem. Theory
  Comput.}\ }\textbf {\bibinfo {volume} {13}},\ \bibinfo {pages} {5273}
  (\bibinfo {year} {2017})}\BibitemShut {NoStop}%
\bibitem [{\citenamefont {Dash}\ \emph {et~al.}(2018)\citenamefont {Dash},
  \citenamefont {Moroni}, \citenamefont {Scemama},\ and\ \citenamefont
  {Filippi}}]{dash2018}%
  \BibitemOpen
  \bibfield  {author} {\bibinfo {author} {\bibfnamefont {M.}~\bibnamefont
  {Dash}}, \bibinfo {author} {\bibfnamefont {S.}~\bibnamefont {Moroni}},
  \bibinfo {author} {\bibfnamefont {A.}~\bibnamefont {Scemama}}, \ and\
  \bibinfo {author} {\bibfnamefont {C.}~\bibnamefont {Filippi}},\ }\href
  {\doibase 10.1021/acs.jctc.8b00393} {\bibfield  {journal} {\bibinfo
  {journal} {J. Chem. Theory Comput.}\ }\textbf {\bibinfo {volume} {14}},\
  \bibinfo {pages} {4176} (\bibinfo {year} {2018})}\BibitemShut {NoStop}%
\bibitem [{\citenamefont {Coldwell}(1977)}]{coldwell1977}%
  \BibitemOpen
  \bibfield  {author} {\bibinfo {author} {\bibfnamefont {R.~L.}\ \bibnamefont
  {Coldwell}},\ }\href {\doibase 10.1002/qua.560120826} {\bibfield  {journal}
  {\bibinfo  {journal} {International Journal of Quantum Chemistry}\ }\textbf
  {\bibinfo {volume} {12}},\ \bibinfo {pages} {215} (\bibinfo {year}
  {1977})}\BibitemShut {NoStop}%
\bibitem [{\citenamefont {Umrigar}\ \emph {et~al.}(1988)\citenamefont
  {Umrigar}, \citenamefont {Wilson},\ and\ \citenamefont
  {Wilkins}}]{umrigar1988}%
  \BibitemOpen
  \bibfield  {author} {\bibinfo {author} {\bibfnamefont {C.~J.}\ \bibnamefont
  {Umrigar}}, \bibinfo {author} {\bibfnamefont {K.~G.}\ \bibnamefont {Wilson}},
  \ and\ \bibinfo {author} {\bibfnamefont {J.~W.}\ \bibnamefont {Wilkins}},\
  }\href {\doibase 10.1103/PhysRevLett.60.1719} {\bibfield  {journal} {\bibinfo
   {journal} {Phys. Rev. Lett.}\ }\textbf {\bibinfo {volume} {60}},\ \bibinfo
  {pages} {1719} (\bibinfo {year} {1988})}\BibitemShut {NoStop}%
\bibitem [{\citenamefont {Malatesta}\ \emph {et~al.}(1997)\citenamefont
  {Malatesta}, \citenamefont {Fahy},\ and\ \citenamefont
  {Bachelet}}]{malatesta1997}%
  \BibitemOpen
  \bibfield  {author} {\bibinfo {author} {\bibfnamefont {A.}~\bibnamefont
  {Malatesta}}, \bibinfo {author} {\bibfnamefont {S.}~\bibnamefont {Fahy}}, \
  and\ \bibinfo {author} {\bibfnamefont {G.~B.}\ \bibnamefont {Bachelet}},\
  }\href {\doibase 10.1103/PhysRevB.56.12201} {\bibfield  {journal} {\bibinfo
  {journal} {Phys. Rev. B}\ }\textbf {\bibinfo {volume} {56}},\ \bibinfo
  {pages} {12201} (\bibinfo {year} {1997})}\BibitemShut {NoStop}%
\bibitem [{\citenamefont {Kent}\ \emph {et~al.}(1999)\citenamefont {Kent},
  \citenamefont {Needs},\ and\ \citenamefont {Rajagopal}}]{kent1999}%
  \BibitemOpen
  \bibfield  {author} {\bibinfo {author} {\bibfnamefont {P.~R.~C.}\
  \bibnamefont {Kent}}, \bibinfo {author} {\bibfnamefont {R.~J.}\ \bibnamefont
  {Needs}}, \ and\ \bibinfo {author} {\bibfnamefont {G.}~\bibnamefont
  {Rajagopal}},\ }\href {\doibase 10.1103/PhysRevB.59.12344} {\bibfield
  {journal} {\bibinfo  {journal} {Phys. Rev. B}\ }\textbf {\bibinfo {volume}
  {59}},\ \bibinfo {pages} {12344} (\bibinfo {year} {1999})}\BibitemShut
  {NoStop}%
\bibitem [{\citenamefont {Umrigar}\ and\ \citenamefont
  {Filippi}(2005)}]{umrigar2005}%
  \BibitemOpen
  \bibfield  {author} {\bibinfo {author} {\bibfnamefont {C.~J.}\ \bibnamefont
  {Umrigar}}\ and\ \bibinfo {author} {\bibfnamefont {C.}~\bibnamefont
  {Filippi}},\ }\href {\doibase 10.1103/PhysRevLett.94.150201} {\bibfield
  {journal} {\bibinfo  {journal} {Phys. Rev. Lett.}\ }\textbf {\bibinfo
  {volume} {94}},\ \bibinfo {pages} {150201} (\bibinfo {year}
  {2005})}\BibitemShut {NoStop}%
\bibitem [{\citenamefont {Shea}\ and\ \citenamefont
  {Neuscamman}(2017)}]{shea2017}%
  \BibitemOpen
  \bibfield  {author} {\bibinfo {author} {\bibfnamefont {J.~A.~R.}\
  \bibnamefont {Shea}}\ and\ \bibinfo {author} {\bibfnamefont {E.}~\bibnamefont
  {Neuscamman}},\ }\href {\doibase 10.1021/acs.jctc.7b00923} {\bibfield
  {journal} {\bibinfo  {journal} {Journal of Chemical Theory and Computation}\
  }\textbf {\bibinfo {volume} {13}},\ \bibinfo {pages} {6078} (\bibinfo {year}
  {2017})}\BibitemShut {NoStop}%
\bibitem [{\citenamefont {Pineda~Flores}\ and\ \citenamefont
  {Neuscamman}(2019)}]{pineda2019}%
  \BibitemOpen
  \bibfield  {author} {\bibinfo {author} {\bibfnamefont {S.~D.}\ \bibnamefont
  {Pineda~Flores}}\ and\ \bibinfo {author} {\bibfnamefont {E.}~\bibnamefont
  {Neuscamman}},\ }\href@noop {} {\bibfield  {journal} {\bibinfo  {journal} {J.
  Phys. Chem. A}\ }\textbf {\bibinfo {volume} {123}},\ \bibinfo {pages} {1487}
  (\bibinfo {year} {2019})}\BibitemShut {NoStop}%
\bibitem [{\citenamefont {Valsson}\ \emph {et~al.}(2012)\citenamefont
  {Valsson}, \citenamefont {Angeli},\ and\ \citenamefont
  {Filippi}}]{valsson2012}%
  \BibitemOpen
  \bibfield  {author} {\bibinfo {author} {\bibfnamefont {O.}~\bibnamefont
  {Valsson}}, \bibinfo {author} {\bibfnamefont {C.}~\bibnamefont {Angeli}}, \
  and\ \bibinfo {author} {\bibfnamefont {C.}~\bibnamefont {Filippi}},\
  }\href@noop {} {\bibfield  {journal} {\bibinfo  {journal} {Phys. Chem. Chem.
  Phys.}\ }\textbf {\bibinfo {volume} {14}},\ \bibinfo {pages} {11015}
  (\bibinfo {year} {2012})}\BibitemShut {NoStop}%
\bibitem [{\citenamefont {Huix-Rotllant}\ \emph {et~al.}(2013)\citenamefont
  {Huix-Rotllant}, \citenamefont {Filatov}, \citenamefont {Gozem},
  \citenamefont {Schapiro}, \citenamefont {Olivucci},\ and\ \citenamefont
  {Ferr\'{e}}}]{huix2013}%
  \BibitemOpen
  \bibfield  {author} {\bibinfo {author} {\bibfnamefont {M.}~\bibnamefont
  {Huix-Rotllant}}, \bibinfo {author} {\bibfnamefont {M.}~\bibnamefont
  {Filatov}}, \bibinfo {author} {\bibfnamefont {S.}~\bibnamefont {Gozem}},
  \bibinfo {author} {\bibfnamefont {I.}~\bibnamefont {Schapiro}}, \bibinfo
  {author} {\bibfnamefont {M.}~\bibnamefont {Olivucci}}, \ and\ \bibinfo
  {author} {\bibfnamefont {N.}~\bibnamefont {Ferr\'{e}}},\ }\href {\doibase
  10.1021/ct4003465} {\bibfield  {journal} {\bibinfo  {journal} {Journal of
  Chemical Theory and Computation}\ }\textbf {\bibinfo {volume} {9}},\ \bibinfo
  {pages} {3917} (\bibinfo {year} {2013})}\BibitemShut {NoStop}%
\bibitem [{\citenamefont {Tuna}\ \emph {et~al.}(2015)\citenamefont {Tuna},
  \citenamefont {Lefrancois}, \citenamefont {Wola\'{n}ki}, \citenamefont
  {Gozem}, \citenamefont {Schapiro}, \citenamefont {Andruni\'{o}w},
  \citenamefont {Dreuw},\ and\ \citenamefont {Olivucci}}]{tuna2015}%
  \BibitemOpen
  \bibfield  {author} {\bibinfo {author} {\bibfnamefont {D.}~\bibnamefont
  {Tuna}}, \bibinfo {author} {\bibfnamefont {D.}~\bibnamefont {Lefrancois}},
  \bibinfo {author} {\bibfnamefont {L.}~\bibnamefont {Wola\'{n}ki}}, \bibinfo
  {author} {\bibfnamefont {S.}~\bibnamefont {Gozem}}, \bibinfo {author}
  {\bibfnamefont {I.}~\bibnamefont {Schapiro}}, \bibinfo {author}
  {\bibfnamefont {T.}~\bibnamefont {Andruni\'{o}w}}, \bibinfo {author}
  {\bibfnamefont {A.}~\bibnamefont {Dreuw}}, \ and\ \bibinfo {author}
  {\bibfnamefont {M.}~\bibnamefont {Olivucci}},\ }\href {\doibase
  10.1021/acs.jctc.5b00022} {\bibfield  {journal} {\bibinfo  {journal} {Journal
  of Chemical Theory and Computation}\ }\textbf {\bibinfo {volume} {11}},\
  \bibinfo {pages} {5758} (\bibinfo {year} {2015})}\BibitemShut {NoStop}%
\bibitem [{\citenamefont {Le~Guennic}\ and\ \citenamefont
  {Jacquemin}(2015)}]{Guennic2015}%
  \BibitemOpen
  \bibfield  {author} {\bibinfo {author} {\bibfnamefont {B.}~\bibnamefont
  {Le~Guennic}}\ and\ \bibinfo {author} {\bibfnamefont {D.}~\bibnamefont
  {Jacquemin}},\ }\href {\doibase 10.1021/ar500447q} {\bibfield  {journal}
  {\bibinfo  {journal} {Accounts of Chemical Research}\ }\textbf {\bibinfo
  {volume} {48}},\ \bibinfo {pages} {530} (\bibinfo {year} {2015})}\BibitemShut
  {NoStop}%
\bibitem [{Jas()}]{Jastrow}%
  \BibitemOpen
  \href@noop {} {}\bibinfo {note} {As Jastrow factor, we use the exponential of
  the sum of two fifth-order polynomials of the electron-nuclear and the
  electron-electron distances, respectively, and rescale the inter-particle
  distances as $R=(1-\exp(-\kappa r))/\kappa$ with $\kappa$ set to 0.6 a.u. We
  employ different electron-nucleus Jastrow factors to describe the correlation
  of an elecron with C and H. The total number of free parameters to be
  optimized in the Jastrow factor is 17 for the systems considered
  here.}\BibitemShut {Stop}%
\bibitem [{\citenamefont {Neuscamman}(2016)}]{neuscamman2016}%
  \BibitemOpen
  \bibfield  {author} {\bibinfo {author} {\bibfnamefont {E.}~\bibnamefont
  {Neuscamman}},\ }\href {\doibase 10.1063/1.4961686} {\bibfield  {journal}
  {\bibinfo  {journal} {The Journal of Chemical Physics}\ }\textbf {\bibinfo
  {volume} {145}},\ \bibinfo {pages} {081103} (\bibinfo {year}
  {2016})}\BibitemShut {NoStop}%
\bibitem [{\citenamefont {Sorella}\ \emph {et~al.}(2007)\citenamefont
  {Sorella}, \citenamefont {Casula},\ and\ \citenamefont
  {Rocca}}]{sorella2007}%
  \BibitemOpen
  \bibfield  {author} {\bibinfo {author} {\bibfnamefont {S.}~\bibnamefont
  {Sorella}}, \bibinfo {author} {\bibfnamefont {M.}~\bibnamefont {Casula}}, \
  and\ \bibinfo {author} {\bibfnamefont {D.}~\bibnamefont {Rocca}},\
  }\href@noop {} {\bibfield  {journal} {\bibinfo  {journal} {J. Chem. Phys.}\
  }\textbf {\bibinfo {volume} {127}},\ \bibinfo {pages} {014105} (\bibinfo
  {year} {2007})}\BibitemShut {NoStop}%
\bibitem [{\citenamefont {Sabzevari}\ \emph {et~al.}(2020)\citenamefont
  {Sabzevari}, \citenamefont {Mahajan},\ and\ \citenamefont
  {Sharma}}]{Sabzevari2020}%
  \BibitemOpen
  \bibfield  {author} {\bibinfo {author} {\bibfnamefont {I.}~\bibnamefont
  {Sabzevari}}, \bibinfo {author} {\bibfnamefont {A.}~\bibnamefont {Mahajan}},
  \ and\ \bibinfo {author} {\bibfnamefont {S.}~\bibnamefont {Sharma}},\
  }\href@noop {} {\bibfield  {journal} {\bibinfo  {journal} {The Journal of
  Chemical Physics}\ }\textbf {\bibinfo {volume} {152}},\ \bibinfo {pages}
  {024111} (\bibinfo {year} {2020})}\BibitemShut {NoStop}%
\bibitem [{\citenamefont {William H.~Press}\ and\ \citenamefont
  {Flannery}(2007)}]{numrec}%
  \BibitemOpen
  \bibfield  {author} {\bibinfo {author} {\bibfnamefont {W.~T.~V.}\
  \bibnamefont {William H.~Press}, \bibfnamefont {Saul A.~Teukolsky}}\ and\
  \bibinfo {author} {\bibfnamefont {B.~P.}\ \bibnamefont {Flannery}},\
  }\href@noop {} {\emph {\bibinfo {title} {Numerical Recepies Third Edition}}}\
  (\bibinfo  {publisher} {Cambridge University Press},\ \bibinfo {address} {The
  Edinburgh Building, Cambridge CB2 8RU, UK},\ \bibinfo {year}
  {2007})\BibitemShut {NoStop}%
\bibitem [{\citenamefont {Toulouse}\ and\ \citenamefont
  {Umrigar}(2008)}]{toulouse2008}%
  \BibitemOpen
  \bibfield  {author} {\bibinfo {author} {\bibfnamefont {J.}~\bibnamefont
  {Toulouse}}\ and\ \bibinfo {author} {\bibfnamefont {C.~J.}\ \bibnamefont
  {Umrigar}},\ }\href {\doibase 10.1063/1.2908237} {\bibfield  {journal}
  {\bibinfo  {journal} {The Journal of Chemical Physics}\ }\textbf {\bibinfo
  {volume} {128}},\ \bibinfo {pages} {174101} (\bibinfo {year}
  {2008})}\BibitemShut {NoStop}%
\bibitem [{Cha()}]{Champ}%
  \BibitemOpen
  \href@noop {} {}\bibinfo {note} {CHAMP is a quantum Monte Carlo program
  package written by C. J. Umrigar, C. Filippi, S. Moroni and
  collaborators}\BibitemShut {NoStop}%
\bibitem [{\citenamefont {Burkatzki}\ \emph {et~al.}(2007)\citenamefont
  {Burkatzki}, \citenamefont {Filippi},\ and\ \citenamefont
  {Dolg}}]{burkatzki2007}%
  \BibitemOpen
  \bibfield  {author} {\bibinfo {author} {\bibfnamefont {M.}~\bibnamefont
  {Burkatzki}}, \bibinfo {author} {\bibfnamefont {C.}~\bibnamefont {Filippi}},
  \ and\ \bibinfo {author} {\bibfnamefont {M.}~\bibnamefont {Dolg}},\
  }\href@noop {} {\bibfield  {journal} {\bibinfo  {journal} {J. Chem. Phys.}\
  }\textbf {\bibinfo {volume} {126}},\ \bibinfo {pages} {234105} (\bibinfo
  {year} {2007})}\BibitemShut {NoStop}%
\bibitem [{BFD()}]{BFD_H2013}%
  \BibitemOpen
  \href@noop {} {}\bibinfo {note} {For the hydrogen atom, we use a more
  accurate BFD pseudopotential and basis set. Dolg, M.; Filippi, C., private
  communication}\BibitemShut {NoStop}%
\bibitem [{\citenamefont {Kendall}\ \emph {et~al.}(1992)\citenamefont
  {Kendall}, \citenamefont {Dunning~Jr},\ and\ \citenamefont
  {Harrison}}]{kendall1992}%
  \BibitemOpen
  \bibfield  {author} {\bibinfo {author} {\bibfnamefont {R.~A.}\ \bibnamefont
  {Kendall}}, \bibinfo {author} {\bibfnamefont {T.~H.}\ \bibnamefont
  {Dunning~Jr}}, \ and\ \bibinfo {author} {\bibfnamefont {R.~J.}\ \bibnamefont
  {Harrison}},\ }\href@noop {} {\bibfield  {journal} {\bibinfo  {journal} {J.
  Chem. Phys.}\ }\textbf {\bibinfo {volume} {96}},\ \bibinfo {pages} {6796}
  (\bibinfo {year} {1992})}\BibitemShut {NoStop}%
\bibitem [{\citenamefont {Attaccalite}\ and\ \citenamefont
  {Sorella}(2008)}]{attaccalite2008}%
  \BibitemOpen
  \bibfield  {author} {\bibinfo {author} {\bibfnamefont {C.}~\bibnamefont
  {Attaccalite}}\ and\ \bibinfo {author} {\bibfnamefont {S.}~\bibnamefont
  {Sorella}},\ }\href@noop {} {\bibfield  {journal} {\bibinfo  {journal} {Phys.
  Rev. Lett.}\ }\textbf {\bibinfo {volume} {100}},\ \bibinfo {pages} {114501}
  (\bibinfo {year} {2008})}\BibitemShut {NoStop}%
\bibitem [{\citenamefont {Casula}(2006)}]{casula2006a}%
  \BibitemOpen
  \bibfield  {author} {\bibinfo {author} {\bibfnamefont {M.}~\bibnamefont
  {Casula}},\ }\href {\doibase 10.1103/PhysRevB.74.161102} {\bibfield
  {journal} {\bibinfo  {journal} {Phys. Rev. B}\ }\textbf {\bibinfo {volume}
  {74}},\ \bibinfo {eid} {161102} (\bibinfo {year} {2006})}\BibitemShut
  {NoStop}%
\bibitem [{\citenamefont {Schmidt}\ \emph {et~al.}(1993)\citenamefont
  {Schmidt}, \citenamefont {Baldridge}, \citenamefont {Boatz}, \citenamefont
  {Elbert}, \citenamefont {Gordon}, \citenamefont {Jensen}, \citenamefont
  {Koseki}, \citenamefont {Matsunaga}, \citenamefont {Nguyen}, \citenamefont
  {Su},\ and\ \citenamefont {{others}}}]{schmidt1993}%
  \BibitemOpen
  \bibfield  {author} {\bibinfo {author} {\bibfnamefont {M.~W.}\ \bibnamefont
  {Schmidt}}, \bibinfo {author} {\bibfnamefont {K.~K.}\ \bibnamefont
  {Baldridge}}, \bibinfo {author} {\bibfnamefont {J.~A.}\ \bibnamefont
  {Boatz}}, \bibinfo {author} {\bibfnamefont {S.~T.}\ \bibnamefont {Elbert}},
  \bibinfo {author} {\bibfnamefont {M.~S.}\ \bibnamefont {Gordon}}, \bibinfo
  {author} {\bibfnamefont {J.~H.}\ \bibnamefont {Jensen}}, \bibinfo {author}
  {\bibfnamefont {S.}~\bibnamefont {Koseki}}, \bibinfo {author} {\bibfnamefont
  {N.}~\bibnamefont {Matsunaga}}, \bibinfo {author} {\bibfnamefont {K.~A.}\
  \bibnamefont {Nguyen}}, \bibinfo {author} {\bibfnamefont {S.}~\bibnamefont
  {Su}}, \ and\ \bibinfo {author} {\bibnamefont {{others}}},\ }\href@noop {}
  {\bibfield  {journal} {\bibinfo  {journal} {J. Comput. Chem.}\ }\textbf
  {\bibinfo {volume} {14}},\ \bibinfo {pages} {1347} (\bibinfo {year}
  {1993})}\BibitemShut {NoStop}%
\bibitem [{\citenamefont {Gordon}\ and\ \citenamefont
  {Schmidt}(2005)}]{gordon2005}%
  \BibitemOpen
  \bibfield  {author} {\bibinfo {author} {\bibfnamefont {M.~S.}\ \bibnamefont
  {Gordon}}\ and\ \bibinfo {author} {\bibfnamefont {M.~W.}\ \bibnamefont
  {Schmidt}},\ }in\ \href@noop {} {\emph {\bibinfo {booktitle} {Theory and
  applications of computational chemistry}}}\ (\bibinfo  {publisher}
  {Elsevier},\ \bibinfo {year} {2005})\ pp.\ \bibinfo {pages}
  {1167--1189}\BibitemShut {NoStop}%
\bibitem [{\citenamefont {Garniron}\ \emph {et~al.}(2019)\citenamefont
  {Garniron}, \citenamefont {Applencourt}, \citenamefont {Gasperich},
  \citenamefont {Benali}, \citenamefont {Fert\'{e}}, \citenamefont {Paquier},
  \citenamefont {Pradines}, \citenamefont {Assaraf}, \citenamefont {Reinhardt},
  \citenamefont {Toulouse}, \citenamefont {Barbaresco}, \citenamefont {Renon},
  \citenamefont {David}, \citenamefont {Malrieu}, \citenamefont {V\'{e}ril},
  \citenamefont {Caffarel}, \citenamefont {Loos}, \citenamefont {Giner},\ and\
  \citenamefont {Scemama}}]{scemama2019}%
  \BibitemOpen
  \bibfield  {author} {\bibinfo {author} {\bibfnamefont {Y.}~\bibnamefont
  {Garniron}}, \bibinfo {author} {\bibfnamefont {T.}~\bibnamefont
  {Applencourt}}, \bibinfo {author} {\bibfnamefont {K.}~\bibnamefont
  {Gasperich}}, \bibinfo {author} {\bibfnamefont {A.}~\bibnamefont {Benali}},
  \bibinfo {author} {\bibfnamefont {A.}~\bibnamefont {Fert\'{e}}}, \bibinfo
  {author} {\bibfnamefont {J.}~\bibnamefont {Paquier}}, \bibinfo {author}
  {\bibfnamefont {B.}~\bibnamefont {Pradines}}, \bibinfo {author}
  {\bibfnamefont {R.}~\bibnamefont {Assaraf}}, \bibinfo {author} {\bibfnamefont
  {P.}~\bibnamefont {Reinhardt}}, \bibinfo {author} {\bibfnamefont
  {J.}~\bibnamefont {Toulouse}}, \bibinfo {author} {\bibfnamefont
  {P.}~\bibnamefont {Barbaresco}}, \bibinfo {author} {\bibfnamefont
  {N.}~\bibnamefont {Renon}}, \bibinfo {author} {\bibfnamefont
  {G.}~\bibnamefont {David}}, \bibinfo {author} {\bibfnamefont {J.-P.}\
  \bibnamefont {Malrieu}}, \bibinfo {author} {\bibfnamefont {M.}~\bibnamefont
  {V\'{e}ril}}, \bibinfo {author} {\bibfnamefont {M.}~\bibnamefont {Caffarel}},
  \bibinfo {author} {\bibfnamefont {P.-F.}\ \bibnamefont {Loos}}, \bibinfo
  {author} {\bibfnamefont {E.}~\bibnamefont {Giner}}, \ and\ \bibinfo {author}
  {\bibfnamefont {A.}~\bibnamefont {Scemama}},\ }\href {\doibase
  10.1021/acs.jctc.9b00176} {\bibfield  {journal} {\bibinfo  {journal} {Journal
  of Chemical Theory and Computation}\ }\textbf {\bibinfo {volume} {15}},\
  \bibinfo {pages} {3591} (\bibinfo {year} {2019})}\BibitemShut {NoStop}%
\bibitem [{das()}]{dash2020}%
  \BibitemOpen
  \href@noop {} {}\bibinfo {note} {Dash, M.; Scemama, A., private
  communication}\BibitemShut {NoStop}%
\bibitem [{\citenamefont {Boulanger}\ \emph {et~al.}(2014)\citenamefont
  {Boulanger}, \citenamefont {Jacquemin}, \citenamefont {Duchemin},\ and\
  \citenamefont {Blase}}]{boulanger2014}%
  \BibitemOpen
  \bibfield  {author} {\bibinfo {author} {\bibfnamefont {P.}~\bibnamefont
  {Boulanger}}, \bibinfo {author} {\bibfnamefont {D.}~\bibnamefont
  {Jacquemin}}, \bibinfo {author} {\bibfnamefont {I.}~\bibnamefont {Duchemin}},
  \ and\ \bibinfo {author} {\bibfnamefont {X.}~\bibnamefont {Blase}},\ }\href
  {\doibase dx.doi.org/10.1021/ct401101u} {\bibfield  {journal} {\bibinfo
  {journal} {J. Chem. Theory Comput.}\ }\textbf {\bibinfo {volume} {10}},\
  \bibinfo {pages} {1212} (\bibinfo {year} {2014})}\BibitemShut {NoStop}%
\bibitem [{\citenamefont {Frisch}\ \emph {et~al.}(2016)\citenamefont {Frisch},
  \citenamefont {Trucks}, \citenamefont {Schlegel}, \citenamefont {Scuseria},
  \citenamefont {Robb}, \citenamefont {Cheeseman}, \citenamefont {Scalmani},
  \citenamefont {Barone}, \citenamefont {Petersson}, \citenamefont {Nakatsuji},
  \citenamefont {Li}, \citenamefont {Caricato}, \citenamefont {Marenich},
  \citenamefont {Bloino}, \citenamefont {Janesko}, \citenamefont {Gomperts},
  \citenamefont {Mennucci}, \citenamefont {Hratchian}, \citenamefont {Ortiz},
  \citenamefont {Izmaylov}, \citenamefont {Sonnenberg}, \citenamefont
  {Williams-Young}, \citenamefont {Ding}, \citenamefont {Lipparini},
  \citenamefont {Egidi}, \citenamefont {Goings}, \citenamefont {Peng},
  \citenamefont {Petrone}, \citenamefont {Henderson}, \citenamefont
  {Ranasinghe}, \citenamefont {Zakrzewski}, \citenamefont {Gao}, \citenamefont
  {Rega}, \citenamefont {Zheng}, \citenamefont {Liang}, \citenamefont {Hada},
  \citenamefont {Ehara}, \citenamefont {Toyota}, \citenamefont {Fukuda},
  \citenamefont {Hasegawa}, \citenamefont {Ishida}, \citenamefont {Nakajima},
  \citenamefont {Honda}, \citenamefont {Kitao}, \citenamefont {Nakai},
  \citenamefont {Vreven}, \citenamefont {Throssell}, \citenamefont
  {Montgomery}, \citenamefont {Peralta}, \citenamefont {Ogliaro}, \citenamefont
  {Bearpark}, \citenamefont {Heyd}, \citenamefont {Brothers}, \citenamefont
  {Kudin}, \citenamefont {Staroverov}, \citenamefont {Keith}, \citenamefont
  {Kobayashi}, \citenamefont {Normand}, \citenamefont {Raghavachari},
  \citenamefont {Rendell}, \citenamefont {Burant}, \citenamefont {Iyengar},
  \citenamefont {Tomasi}, \citenamefont {Cossi}, \citenamefont {Millam},
  \citenamefont {Klene}, \citenamefont {Adamo}, \citenamefont {Cammi},
  \citenamefont {Ochterski}, \citenamefont {Martin}, \citenamefont {Morokuma},
  \citenamefont {Farkas}, \citenamefont {Foresman},\ and\ \citenamefont
  {Fox.}}]{Gaussian09}%
  \BibitemOpen
  \bibfield  {author} {\bibinfo {author} {\bibfnamefont {M.~J.}\ \bibnamefont
  {Frisch}}, \bibinfo {author} {\bibfnamefont {G.~W.}\ \bibnamefont {Trucks}},
  \bibinfo {author} {\bibfnamefont {H.~B.}\ \bibnamefont {Schlegel}}, \bibinfo
  {author} {\bibfnamefont {G.~E.}\ \bibnamefont {Scuseria}}, \bibinfo {author}
  {\bibfnamefont {M.~A.}\ \bibnamefont {Robb}}, \bibinfo {author}
  {\bibfnamefont {J.~R.}\ \bibnamefont {Cheeseman}}, \bibinfo {author}
  {\bibfnamefont {G.}~\bibnamefont {Scalmani}}, \bibinfo {author}
  {\bibfnamefont {V.}~\bibnamefont {Barone}}, \bibinfo {author} {\bibfnamefont
  {G.~A.}\ \bibnamefont {Petersson}}, \bibinfo {author} {\bibfnamefont
  {H.}~\bibnamefont {Nakatsuji}}, \bibinfo {author} {\bibfnamefont
  {X.}~\bibnamefont {Li}}, \bibinfo {author} {\bibfnamefont {M.}~\bibnamefont
  {Caricato}}, \bibinfo {author} {\bibfnamefont {A.}~\bibnamefont {Marenich}},
  \bibinfo {author} {\bibfnamefont {J.}~\bibnamefont {Bloino}}, \bibinfo
  {author} {\bibfnamefont {B.~G.}\ \bibnamefont {Janesko}}, \bibinfo {author}
  {\bibfnamefont {R.}~\bibnamefont {Gomperts}}, \bibinfo {author}
  {\bibfnamefont {B.}~\bibnamefont {Mennucci}}, \bibinfo {author}
  {\bibfnamefont {H.~P.}\ \bibnamefont {Hratchian}}, \bibinfo {author}
  {\bibfnamefont {J.~V.}\ \bibnamefont {Ortiz}}, \bibinfo {author}
  {\bibfnamefont {A.~F.}\ \bibnamefont {Izmaylov}}, \bibinfo {author}
  {\bibfnamefont {J.~L.}\ \bibnamefont {Sonnenberg}}, \bibinfo {author}
  {\bibfnamefont {D.}~\bibnamefont {Williams-Young}}, \bibinfo {author}
  {\bibfnamefont {F.}~\bibnamefont {Ding}}, \bibinfo {author} {\bibfnamefont
  {F.}~\bibnamefont {Lipparini}}, \bibinfo {author} {\bibfnamefont
  {F.}~\bibnamefont {Egidi}}, \bibinfo {author} {\bibfnamefont
  {J.}~\bibnamefont {Goings}}, \bibinfo {author} {\bibfnamefont
  {B.}~\bibnamefont {Peng}}, \bibinfo {author} {\bibfnamefont {A.}~\bibnamefont
  {Petrone}}, \bibinfo {author} {\bibfnamefont {T.}~\bibnamefont {Henderson}},
  \bibinfo {author} {\bibfnamefont {D.}~\bibnamefont {Ranasinghe}}, \bibinfo
  {author} {\bibfnamefont {V.~G.}\ \bibnamefont {Zakrzewski}}, \bibinfo
  {author} {\bibfnamefont {J.}~\bibnamefont {Gao}}, \bibinfo {author}
  {\bibfnamefont {N.}~\bibnamefont {Rega}}, \bibinfo {author} {\bibfnamefont
  {G.}~\bibnamefont {Zheng}}, \bibinfo {author} {\bibfnamefont
  {W.}~\bibnamefont {Liang}}, \bibinfo {author} {\bibfnamefont
  {M.}~\bibnamefont {Hada}}, \bibinfo {author} {\bibfnamefont {M.}~\bibnamefont
  {Ehara}}, \bibinfo {author} {\bibfnamefont {K.}~\bibnamefont {Toyota}},
  \bibinfo {author} {\bibfnamefont {R.}~\bibnamefont {Fukuda}}, \bibinfo
  {author} {\bibfnamefont {J.}~\bibnamefont {Hasegawa}}, \bibinfo {author}
  {\bibfnamefont {M.}~\bibnamefont {Ishida}}, \bibinfo {author} {\bibfnamefont
  {T.}~\bibnamefont {Nakajima}}, \bibinfo {author} {\bibfnamefont
  {Y.}~\bibnamefont {Honda}}, \bibinfo {author} {\bibfnamefont
  {O.}~\bibnamefont {Kitao}}, \bibinfo {author} {\bibfnamefont
  {H.}~\bibnamefont {Nakai}}, \bibinfo {author} {\bibfnamefont
  {T.}~\bibnamefont {Vreven}}, \bibinfo {author} {\bibfnamefont
  {K.}~\bibnamefont {Throssell}}, \bibinfo {author} {\bibfnamefont {J.~A.}\
  \bibnamefont {Montgomery}}, \bibinfo {author} {\bibfnamefont {J.~J.~E.}\
  \bibnamefont {Peralta}}, \bibinfo {author} {\bibfnamefont {F.}~\bibnamefont
  {Ogliaro}}, \bibinfo {author} {\bibfnamefont {M.}~\bibnamefont {Bearpark}},
  \bibinfo {author} {\bibfnamefont {J.~J.}\ \bibnamefont {Heyd}}, \bibinfo
  {author} {\bibfnamefont {E.}~\bibnamefont {Brothers}}, \bibinfo {author}
  {\bibfnamefont {K.~N.}\ \bibnamefont {Kudin}}, \bibinfo {author}
  {\bibfnamefont {V.~N.}\ \bibnamefont {Staroverov}}, \bibinfo {author}
  {\bibfnamefont {T.}~\bibnamefont {Keith}}, \bibinfo {author} {\bibfnamefont
  {R.}~\bibnamefont {Kobayashi}}, \bibinfo {author} {\bibfnamefont
  {J.}~\bibnamefont {Normand}}, \bibinfo {author} {\bibfnamefont
  {K.}~\bibnamefont {Raghavachari}}, \bibinfo {author} {\bibfnamefont
  {A.}~\bibnamefont {Rendell}}, \bibinfo {author} {\bibfnamefont {J.~C.}\
  \bibnamefont {Burant}}, \bibinfo {author} {\bibfnamefont {S.~S.}\
  \bibnamefont {Iyengar}}, \bibinfo {author} {\bibfnamefont {J.}~\bibnamefont
  {Tomasi}}, \bibinfo {author} {\bibfnamefont {M.}~\bibnamefont {Cossi}},
  \bibinfo {author} {\bibfnamefont {J.~M.}\ \bibnamefont {Millam}}, \bibinfo
  {author} {\bibfnamefont {M.}~\bibnamefont {Klene}}, \bibinfo {author}
  {\bibfnamefont {C.}~\bibnamefont {Adamo}}, \bibinfo {author} {\bibfnamefont
  {R.}~\bibnamefont {Cammi}}, \bibinfo {author} {\bibfnamefont {J.~W.}\
  \bibnamefont {Ochterski}}, \bibinfo {author} {\bibfnamefont {R.~L.}\
  \bibnamefont {Martin}}, \bibinfo {author} {\bibfnamefont {K.}~\bibnamefont
  {Morokuma}}, \bibinfo {author} {\bibfnamefont {O.}~\bibnamefont {Farkas}},
  \bibinfo {author} {\bibfnamefont {J.~B.}\ \bibnamefont {Foresman}}, \ and\
  \bibinfo {author} {\bibfnamefont {D.~J.}\ \bibnamefont {Fox.}},\ }\href@noop
  {} {\enquote {\bibinfo {title} {Gaussian09, revision a.02; gaussian, inc.,
  wallingford ct},}\ } (\bibinfo {year} {2016})\BibitemShut {NoStop}%
\bibitem [{\citenamefont {Parrish}\ \emph {et~al.}(2017)\citenamefont
  {Parrish}, \citenamefont {Burns}, \citenamefont {Smith}, \citenamefont
  {Simmonett}, \citenamefont {DePrince~III}, \citenamefont {Hohenstein},
  \citenamefont {Bozkaya}, \citenamefont {Sokolov}, \citenamefont {Di~Remigio},
  \citenamefont {Richard},\ and\ \citenamefont {{others}}}]{psi4}%
  \BibitemOpen
  \bibfield  {author} {\bibinfo {author} {\bibfnamefont {R.~M.}\ \bibnamefont
  {Parrish}}, \bibinfo {author} {\bibfnamefont {L.~A.}\ \bibnamefont {Burns}},
  \bibinfo {author} {\bibfnamefont {D.~G.}\ \bibnamefont {Smith}}, \bibinfo
  {author} {\bibfnamefont {A.~C.}\ \bibnamefont {Simmonett}}, \bibinfo {author}
  {\bibfnamefont {A.~E.}\ \bibnamefont {DePrince~III}}, \bibinfo {author}
  {\bibfnamefont {E.~G.}\ \bibnamefont {Hohenstein}}, \bibinfo {author}
  {\bibfnamefont {U.}~\bibnamefont {Bozkaya}}, \bibinfo {author} {\bibfnamefont
  {A.~Y.}\ \bibnamefont {Sokolov}}, \bibinfo {author} {\bibfnamefont
  {R.}~\bibnamefont {Di~Remigio}}, \bibinfo {author} {\bibfnamefont {R.~M.}\
  \bibnamefont {Richard}}, \ and\ \bibinfo {author} {\bibnamefont {{others}}},\
  }\href@noop {} {\bibfield  {journal} {\bibinfo  {journal} {J. Chem. Theory
  Comput.}\ }\textbf {\bibinfo {volume} {13}},\ \bibinfo {pages} {3185}
  (\bibinfo {year} {2017})}\BibitemShut {NoStop}%
\bibitem [{\citenamefont {Garniron}\ \emph {et~al.}(2018)\citenamefont
  {Garniron}, \citenamefont {Scemama}, \citenamefont {Giner}, \citenamefont
  {Caffarel},\ and\ \citenamefont {Loos}}]{garniron2018}%
  \BibitemOpen
  \bibfield  {author} {\bibinfo {author} {\bibfnamefont {Y.}~\bibnamefont
  {Garniron}}, \bibinfo {author} {\bibfnamefont {A.}~\bibnamefont {Scemama}},
  \bibinfo {author} {\bibfnamefont {E.}~\bibnamefont {Giner}}, \bibinfo
  {author} {\bibfnamefont {M.}~\bibnamefont {Caffarel}}, \ and\ \bibinfo
  {author} {\bibfnamefont {P.-F.}\ \bibnamefont {Loos}},\ }\href {\doibase
  10.1063/1.5044503} {\bibfield  {journal} {\bibinfo  {journal} {The Journal of
  Chemical Physics}\ }\textbf {\bibinfo {volume} {149}},\ \bibinfo {pages}
  {064103} (\bibinfo {year} {2018})}\BibitemShut {NoStop}%
\end{thebibliography}%

\end{document}